\documentclass[12pt,preprint]{aastex}

\newcommand{\Msun}{~M_\odot}
\newcommand{\lsim}{\raise0.3ex\hbox{$<$}\kern-0.75em{\lower0.65ex\hbox{$\sim$}}}
\newcommand{\gsim}{\raise0.3ex\hbox{$>$}\kern-0.75em{
\lower0.65ex\hbox{$\sim$}}}

\newcommand{\cmc}{\rm ~cm^{-3}}

\newcommand{\kms}{\rm ~km~s^{-1}}
\newcommand{\ergs}{\rm ~erg~s^{-1}}
\newcommand{\wl}{\lambda}
\newcommand{\wll}{\lambda \lambda}

\newcommand{\Ha}{${\rm H}\alpha$}
\newcommand{\La}{${\rm Ly}\alpha$}
\newcommand{\Hb}{${\rm H}\beta$}

\shorttitle{CNO Processing in Supernova Progenitors}
\shortauthors{Fransson et al.}

\received{2004 June 23}
\begin{document}


\title{Hubble Space Telescope and Ground-Based Observations of SN 1993J 
and SN 1998S: CNO Processing in the Progenitors\altaffilmark{1}}


\author{
Claes Fransson\altaffilmark{2},
Peter M. Challis\altaffilmark{3},
Roger A. Chevalier\altaffilmark{4},
Alexei V. Filippenko\altaffilmark{5},
Robert P. Kirshner\altaffilmark{3},
Cecilia Kozma\altaffilmark{2},
Douglas C. Leonard\altaffilmark{6,7},
Thomas Matheson\altaffilmark{3},
E. Baron\altaffilmark{8},
Peter Garnavich\altaffilmark{9}, 
Saurabh Jha\altaffilmark{5},
Bruno Leibundgut\altaffilmark{10},
Peter Lundqvist\altaffilmark{2},
C. S. J. Pun\altaffilmark{11}, 
Lifan Wang\altaffilmark{12}, and 
J. Craig Wheeler\altaffilmark{13}
}

\altaffiltext{1}{Based in part on observations obtained with the {\it
Hubble Space Telescope,} which is operated by AURA, Inc., under NASA
contract NAS 5-26555.}
\altaffiltext{2}{Department of Astronomy, Stockholm University, AlbaNova, 
SE--106~91 Stockholm, Sweden; claes@astro.su.se email.}
\altaffiltext{3}{Harvard--Smithsonian Center
for Astrophysics, 60 Garden St., Cambridge, MA 02138.}
\altaffiltext{4}{Department of Astronomy, University of
Virginia, P.O. Box 3818, Charlottesville, VA 22903.}
\altaffiltext{5}{Department of Astronomy, University of California,
Berkeley, CA 94720--3411.}
\altaffiltext{6}{Five College Astronomy Department,
University of Massachusetts, Amherst, MA 01003-9305.}
\altaffiltext{7}{Department of Astronomy, 105-24 Caltech, Pasadena,
CA 91125.}
\altaffiltext{8}{Department of Physics and Astronomy, University of Oklahoma,
440 W. Brooks, Norman, OK 73019-0261.}
\altaffiltext{9}{Department of Physics, University of Notre Dame, 225
Nieuwland Science Hall, Notre Dame, IN 45656.}
\altaffiltext{10}{European Southern Observatory,
Karl-Schwarzschild-Strasse 2, D-85748 Garching, Germany.}
\altaffiltext{11}{Department of Physics, University of Hong Kong, Pokfulam
Road, Hong Kong.}
\altaffiltext{12}{Institute for Nuclear and Particle Astrophysics,
E. O. Lawrence Berkeley National Laboratory, Berkeley, CA 94720.}
\altaffiltext{13}{Department of Astronomy, University of Texas, Austin, TX
78712.}


\begin{abstract}

Ground-based and {\it Hubble Space Telescope} observations are
presented for SN 1993J and SN 1998S. SN 1998S shows strong, relatively
narrow circumstellar emission lines of N~III-V and C~III-IV, as well
as broad lines from the ejecta. Both the broad ultraviolet and optical
lines in SN 1998S indicate an expansion velocity of $\sim7,000
\kms$. The broad emission components of Ly$\alpha$ and Mg~II are
strongly asymmetrical after day 72 past the explosion, and differ in
shape from H$\alpha$. Different models based on dust extinction from
dust in the ejecta or shock region, in combination with \Ha\ from a
circumstellar torus, are discussed. It is concluded, however, that the
double-peaked line profiles are more likely to arise as a result of
optical depth effects in the narrow, cool, dense shell behind the
reverse shock, than in a torus-like region. The ultraviolet lines of
SN 1993J are broad, with a box-like shape, coming from the ejecta and
a cool dense shell. The shapes of the lines are well fitted with a
shell with inner velocity $\sim7,000 \kms$ and outer velocity
$\sim10,000 \kms$. For both SN 1993J and SN 1998S a strong nitrogen
enrichment is found, with N/C $\approx 12.4$ in SN 1993J and N/C
$\approx 6.0$ in SN 1998S. From a compilation of all supernovae with
determined CNO ratios, we discuss the implications of these
observations for the structure of the progenitors of Type II
supernovae.

\end{abstract}




\section{INTRODUCTION}
\label{sec_intr}
The nature of the progenitors of the different types of supernovae (SNe)
is still debated. One of the main unknowns is the amount of mass loss
suffered by the progenitor before the explosion. The extent of the
mass loss, driven either by stellar winds in single stars or by binary
evolution, causes widely different progenitor structures. Because the
supernova light curve and spectrum to a large extent are functions of
the mass of the hydrogen and helium envelope, the mass-loss history is
a key ingredient in understanding the various observational
signatures, manifested in the many separate classes of core-collapse
SNe.  Mass loss affects the chemical composition of the ejecta; hence,
detailed spectral studies of the different types of SNe and their
environments can shed some light on this issue.

A large number of SNe have now shown various types and degrees of
circumstellar interaction; see \cite{F97} for a recent review. 
Both of the ``Type II-linear supernovae'' (SNe~II-L)
1979C and 1980K showed evidence for circumstellar interaction in their
radio emission \citep{W86}, as well as in the ultraviolet (UV)
emission in the case of SN 1979C \citep{P80,Fes99}. The circumstellar
medium most likely originates from the dense, slow superwind of the
red supergiant progenitor. Type Ib and Ic SNe are believed to have
lost most of their hydrogen envelope prior to exploding. They in
general show both radio emission and X-ray emission caused by
circumstellar interaction. Because of the high wind speed ($\ga 1,000
\kms$) of the Wolf-Rayet (WR) progenitor, the wind in this case is
considerably less dense compared to that of SNe~II-L and II-P.

SN 1993J provided a link between SNe~II and SNe~Ib in that it
underwent a transition from a Type II spectrum at early epochs to a
Type Ib-like spectrum at $\sim300$ days \citep{FMH93,FMB94,FF95,Bar95}; see
\cite{FM04} for a recent review. At early epochs it already showed the
typical signatures of circumstellar interaction at radio \citep{VD94},
UV \citep{FS94}, and X-ray \citep{Z94a} wavelengths. After about one
year, the optical spectrum became dominated by emission lines excited
by the circumstellar interaction \citep{FMB94,PCM95,MFa00,MFb00}.

Type IIn supernovae (e.g., Filippenko 1997) have optical spectral signatures of
a circumstellar medium (CSM) present already at early phases. Although they
show considerable variation from object to object, the common characteristics
are strong, relatively narrow emission lines after a few weeks, or perhaps
earlier; these lines are thought to be produced by the interaction between the
supernova ejecta and the CSM.  In extreme cases (e.g., SN 1988Z; Turatto et
al. 1993; Chugai \& Danziger 1994) the CSM is clearly very dense and extensive,
and the interaction lasts for many years. Occasionally (e.g., SN 1994W; Chugai
et al. 2004) there is evidence that much of the CSM was produced just a short
time before the explosion.  Also, the light curves of SNe~IIn have a large
variation in terms of the relative importance of the radioactive heating and
energy input from circumstellar interaction. However, the nature and relation
of the SNe~IIn to other SN types are not clear \citep[for a discussion,
see][]{NIS95}.  SN 1998S represents one of the best-studied examples of this
type of supernova.

In contrast to SN 1993J, the optical spectrum of SN 1998S was
dominated by circumstellar interaction even at early
epochs. \cite{BRMB00} found that many low-ionization UV lines had a
narrow P-Cygni component, which they interpret as a wind with velocity
$\sim50 \kms$, as well as a more highly ionized component with
velocity $\sim300 \kms$. This was later confirmed from ground-based
observations by \cite{Fas01}.  \citet{Ch01} argued that the narrow
core and smooth high-velocity wings of the H$\alpha$ line at early
epochs could be understood as a result of electron scattering in a
dense circumstellar shell extending from the supernova. While the
observations during the first months showed fairly symmetrical line
profiles, the lines changed character after $\sim100$ days: H$\alpha$,
as well as He~I $\wl 10830$, displayed a highly asymmetrical
triple-peaked structure \citep{Ger00}. Infrared observations showed
evidence for CO formation at $\sim95$ days, and dust formation at
$\sim225$ days \citep{Fas00,Ger02}. The supernova also showed strong X-ray
emission at late epochs \citep{Poo02}. No radio emission was seen at
early epochs. However, by 310 days emission was detected at 8.46 GHz,
and subsequently also at 1.47 GHz and 4.89 GHz \citep{Poo02}.  The
late turn-on and its frequency dependence may be understood as a
result of strong free-free absorption by the circumstellar gas.

One of the most important indicators of the extent of mass loss is the
relative abundances of the CNO elements. Depending on the mass lost
and the degree of mixing, CNO burning products may be seen in either
the circumstellar medium or the outer parts of the progenitor, and
therefore in the supernova \citep[e.g.,][]{MM00,HL00,WL99}. Evidence
for such CNO processing has earlier been seen in SN 1979C
\citep{FB84}, SN 1987A \citep{FC89}, and SN 1995N \citep{FC02}. In
this paper we add two more cases.

SN 1993J and the Type IIn SN 1998S are among the best-observed
circumstellar interactors to date. In this paper we report on UV
observations of these two SNe, and on their implications for the
abundances of the CNO elements.  Early-epoch {\it Hubble Space
Telescope (HST)} observations of both SNe have previously been
discussed, but mainly in the context of the outer regions of the
supernova ejecta \citep{Jef94,HF96,Len01}. In this paper we
concentrate on the late-epoch observations, where most of the
indicators of circumstellar interaction are present. In addition to
the {\it HST} observations, we also add new ground-based
observations. Based on the complete evolution of the line profiles, we
discuss implications of the shapes of these for the physics of the
interaction region. The analysis in this phase is simplified by the
basically nebular conditions of the line-emitting gas.

SN 1993J was discovered on 1993 March 28.9 \citep{Rip93}, while SN
1998S was discovered on 1998 March 2.68 UT (Li et al. 1998; Qui et
al. 1998). Here we adopt explosion dates of 1993 March 27.5 for SN
1993J \citep{Lew94}, and 1998 March 2 for SN 1998S. The true explosion
date of SN 1998S was probably somewhat earlier, of course, but by a
negligible amount given the late-time phases studied here.  \cite{R94}
discuss the reddening of SN 1993J and argue for a most likely value of
$E_{B-V} = 0.2$ mag, which we adopt.  For SN 1998S \cite{Leo00} find a
reddening of $E_{B-V} = 0.23$ mag.  The distance to SN 1993J is 3.63
Mpc \citep{F94} and the recession velocity is $-135 \kms$
\citep{VC94}. The recession velocity of SN 1998S is $846.9 \kms$
\citep{Fas01}, and we adopt a distance of 17 Mpc \citep{Tul88}.

\section{OBSERVATIONS}

The {\it HST} observations in this paper were obtained with the Faint
Object Spectrograph (FOS) and the Space Telescope Imaging Spectrograph
(STIS). Tables \ref{tab1a} and \ref{tab1b} give the journal of
observations of SN 1993J and SN 1998S (respectively), including
exposure times, gratings used, their dispersion, and the wavelength
ranges covered.  The FOS spectra were
obtained with the G160L, G270H, and G400H gratings, while the STIS
spectra were obtained with gratings G140L, G230L, and G430L.  All
spectra were calibrated by the FOS and STIS pipelines at the Space
Telescope Science Institute.  The \La\ region in the FOS spectra of SN
1993J on days 1063 and 1399 is dominated by the geocoronal \La, so it
is not used in this paper.  In the STIS spectra on days 1792--2585
this component could be subtracted successfully.

To complement the {\it HST} observations, optical low-dispersion
spectra of SN 1998S for days 29, 75, 257--258, 431, and 440 were
obtained with the FAST spectrograph \citep{FCC98} on the 1.5-m
Tillinghast telescope at the Fred L. Whipple Observatory (FLWO).  The
FAST spectrograph employs a $2688 \times 512$ pixel Loral CCD with a
spatial scale of 1.$\arcsec$1 pixel$^{-1}$ in the binning mode used
for these observations.  Details of the exposures are given in Table
\ref{tab1c}.  The data were reduced in the standard manner with
IRAF\footnote{IRAF is distributed by the National Optical Astronomy
Observatories, which are operated by the Association of Universities
for Research in Astronomy, Inc., under cooperative agreement with the
National Science Foundation.} and our own routines.  Wavelength
calibration was accomplished with He-Ne-Ar lamps taken immediately
after each supernova exposure.  Small-scale wavelength adjustments
derived from night-sky lines in the supernova frames were also
applied.  Spectrophotometric standards used in the reductions are
listed in Table \ref{tab1c}.  We attempted to remove telluric lines
using the well-exposed continua of the spectrophotometric standards
\citep{MFa00}.  The FLWO spectra were observed, in general, within
$\sim10\degr$ of the parallactic angle to minimize losses from
atmospheric dispersion \citep{Fil82}. 

To complete the temporal coverage of the optical spectra of SN 1998S
we have also included spectra up to day 494 from Lick Observatory,
published by \citet{Leo00}; see Table \ref{tab1c}. In addition to
these, two spectra on days 643 and 653 were also obtained at the
Keck~II 10-m telescope with LRIS \citep{Oke95}, and one very late-time
spectrum (day 2148) was obtained with the MMT 6.5-m telescope using
the Blue Channel spectrograph \citep{Sch89}.  All data were reduced in
a manner similar to that described above for the FLWO spectra. For the
spectra on days 643 and 2148 the conditions were such that a proper
background subtraction was difficult. These spectra may therefore be
contaminated by the galaxy background and the fluxes may be
correspondingly uncertain. Note that our spectrum from day
653 is also shown in the \Ha \ compilation in \citet{Poz04}. 

Details of the Keck observations of SN 1993J on days 670, 976, 1766,
and 2454, included in this paper, are given by \citet{MFa00}. In
addition, we have included one spectrum of SN 1993J taken with MMT
on 1997 April 8 (day 1473).

\section{RESULTS}

\subsection{SN 1993J}
\label{sec_93j}

{\it HST} observations of SN 1993J at early phases were discussed in
\citet{Jef94} and \cite{BHB94}. In this paper we limit ourselves to
the nebular phase, when the spectrum is dominated by the circumstellar
interaction, and the line profiles are easier to identify and analyze.

\begin{figure}
\epsscale{0.85}
\plotone{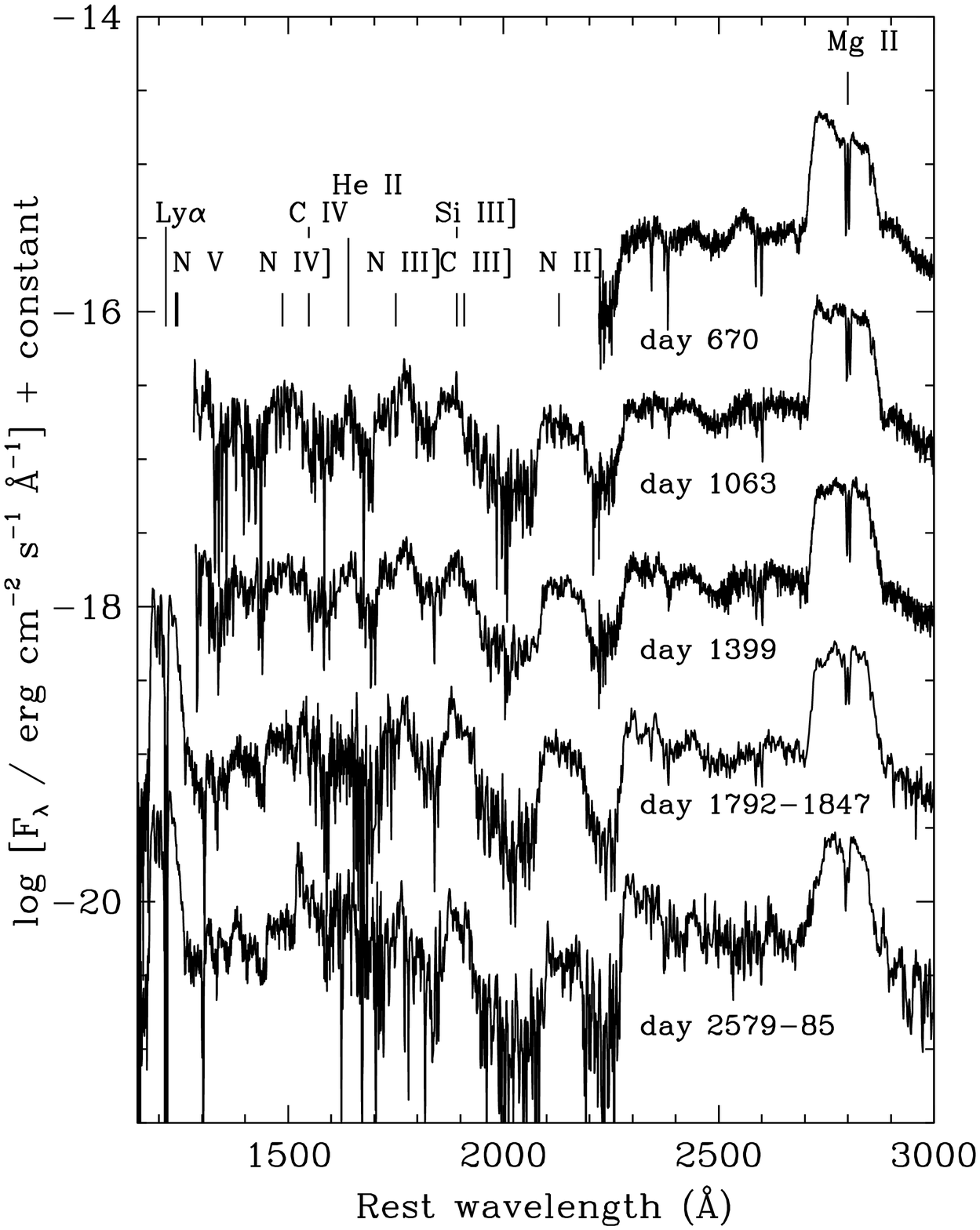}
\caption{The late-time evolution of the spectrum of SN 1993J, with epochs (days
after the explosion) indicated. Each spectrum is shifted down by 1.0 dex with
respect to the previous one.  The \La\ line in the spectra at days 1063 and
1399 is severely affected by geocoronal \La, and has therefore been omitted.
\label{fig00}}
\end{figure}

Figure \ref{fig00} shows the {\it HST} spectra of SN 1993J from
January 1995 to April 2000 (days 670 to 2585), where we have noted the
positions of some of the strongest spectral features. As will be
discussed below, however, some of these are likely to be blends of
several lines.  Note that the overlap region of the short-wavelength
and long-wavelength UV detectors in the range 1600--1700~\AA\ is
affected by an increased noise level.

During this whole time in the nebular phase, the spectrum is
remarkably constant. The far-UV region is dominated by the strong
Ly$\alpha$ emission line, probably blended with N~V $\wll 1238.8,
1242.8$. The region shortward of $\sim2000$~\AA\ forms a
pseudo-continuum of blended lines of highly ionized nitrogen, carbon,
and oxygen, which we discuss in more detail below.  The well-defined
line at $\sim2140$~\AA\ is most likely due to N~II] $\wll 2139.7,
2143.5$. The presence of this prominent line indicates a strong
nitrogen enrichment, as we will confirm later. The plateau longward of
$\sim2280$~\AA\ is probably formed mainly of Fe~II resonance lines,
although a blend of C~II] 2323.5--2328.1, O~III] 2320.9--2331.4, and
Si~II] $\wl 2334.6$ is likely to explain the feature at
$\sim2332$~\AA. Finally, the near-UV region is dominated by the
prominent Mg~II $\wll 2795.5, 2802.7$ doublet.

\begin{figure}
\plotone{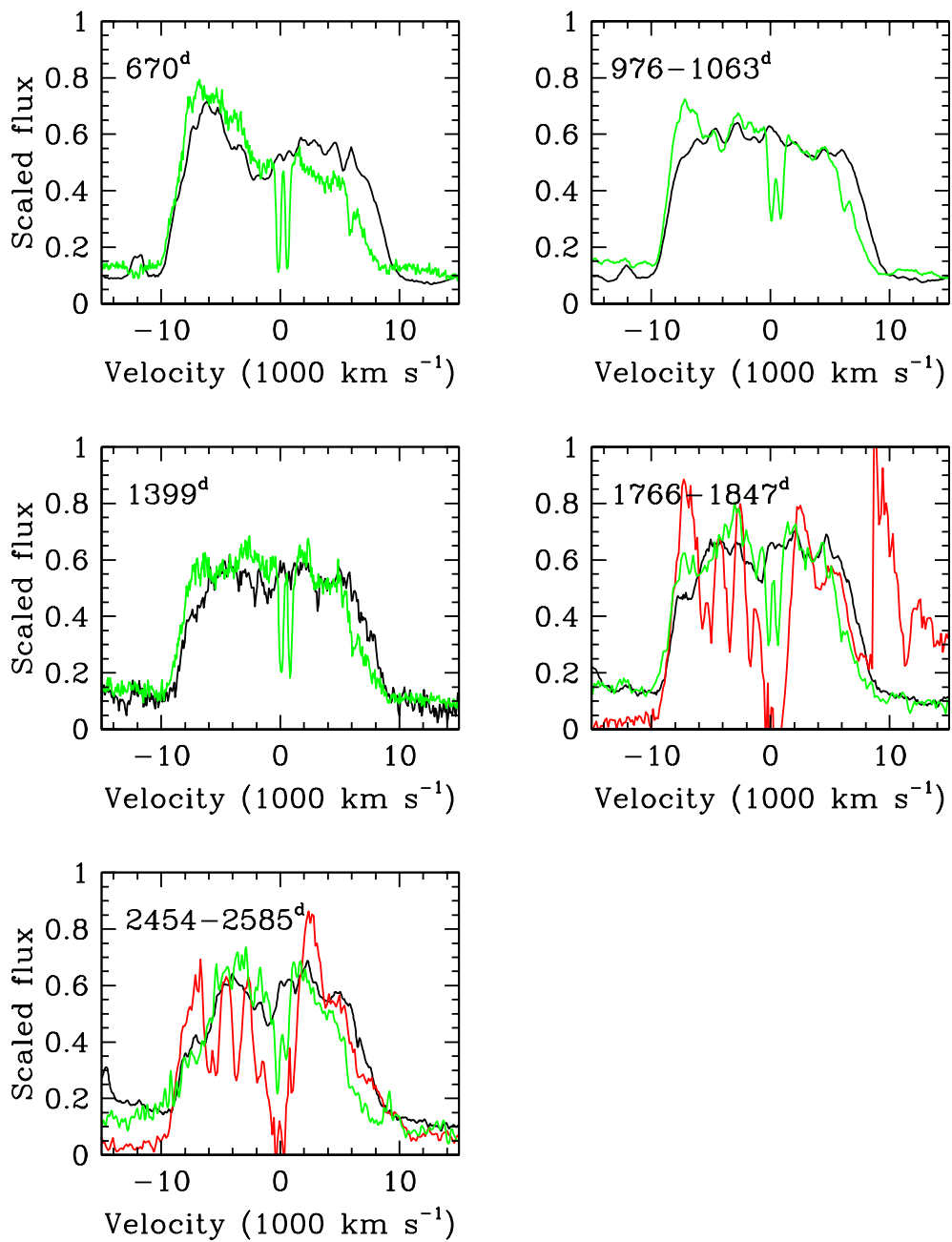}
\caption{Evolution of the line profiles of H$\alpha$ (black), Ly$\alpha$ (red),
and Mg~II (green) in SN 1993J. The flux scale differs among the lines. The
epoch (days after the explosion) is given in each panel. In the {\it HST} spectra
from days 1063 and 1399 the \La\ line is severely contaminated by geocoronal
\La, and is therefore not shown. There was no observation of \La\ on day 670.
\label{fig1a}}
\end{figure}

The most interesting temporal change is in the shape of the Mg~II
line. This evolution is similar to that seen in H$\alpha$
\citep{MFa00}. In Figure \ref{fig1a} we compare the Ly$\alpha$,
H$\alpha$, and Mg~II line profiles.  The H$\alpha$ profiles are taken
from \cite{MFa00}. From the figure we see that the shapes of both the
H$\alpha$ and Mg~II lines change in concordance with each other. On
day 670 both lines are clearly asymmetric, with the blue side of Mg~II
a factor of $\sim2.2$ stronger than the red. The asymmetry of
H$\alpha$ is considerably smaller, with a blue-to-red intensity ratio
of $\sim1.1$. By day 1063 this asymmetry has decreased for both lines,
and for the later epochs the two sides have nearly equal strength. The
Ly$\alpha$ line is strongly affected by geocoronal emission, which
makes a comparison with the other lines uncertain.

As has been previously noted \citep[e.g.,][]{FMB94,FF95,PCM95,MFb00},
the line profiles after $\sim1$ year, in particular that of H$\alpha$,
can be best described as box-like. This strongly indicates that the
emission comes from a relatively thin shell in the expanding
gas. \cite{MFb00} found during this period a maximum velocity of
$\sim9,000$--10,600 $\kms$ for the blue edge of the H$\alpha$ line,
and a slightly lower velocity for the red edge.

Of the UV lines, only the Mg~II and [N~II] $\wll 2139.7, 2143.5$ lines
are unblended and have sufficiently high signal-to-noise ratio (S/N)
to warrant a detailed line fit. As is seen from Figure \ref{fig00},
the UV spectrum does not change appreciably from January 1996 to
February 1997. To increase the S/N, we therefore average the day 1063
and day 1399 spectra. We then fit these line profiles, as well as the
H$\alpha$ profile from the Keck spectrum on day 976, with a shell
having inner velocity $V_{\rm in}$ and outer velocity $V_{\rm out}$;
see \cite{Fra84} for details. The variation of the emissivity within
the shell is of minor importance; as long as the shell is reasonably
thin, it only affects the wings of the lines (for $V > V_{\rm in}$).

\begin{figure}
\epsscale{1.0}
\plotone{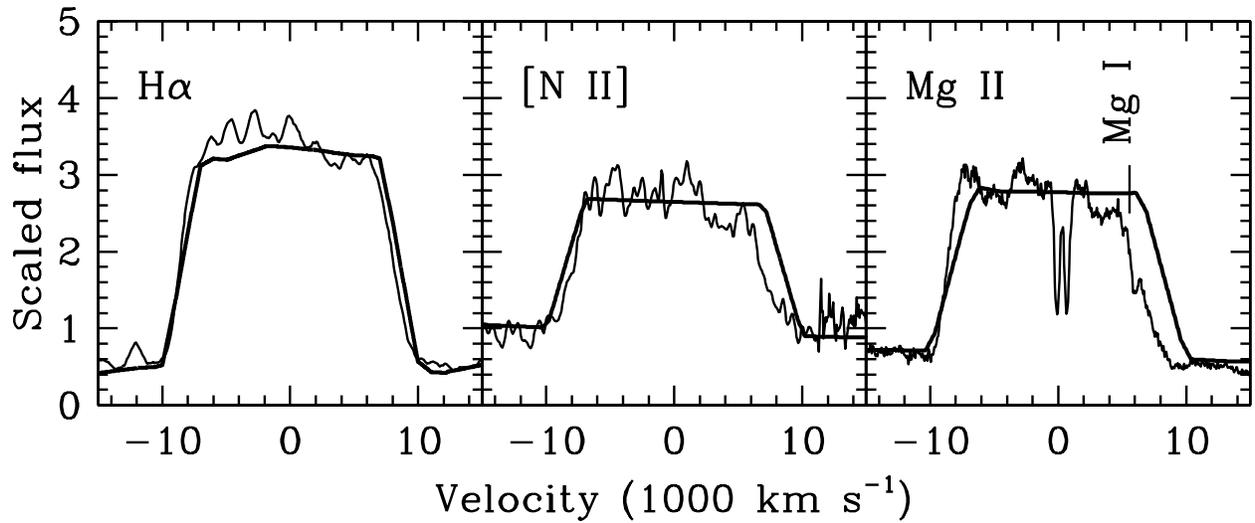}
\caption{Observed line profiles of H$\alpha$, [N~II], and Mg~II in SN
1993J. For [N~II] and Mg~II we have averaged the day 1063 and 1399
spectra, while H$\alpha$ is from the day 976 Keck spectrum. The flux
scale differs among the lines. The thick line gives the expected
line profile for a shell of constant emissivity, inner velocity
$V_{\rm in}=7,000 \kms$, and outer velocity $V_{\rm out}=10,000 \kms$.
\label{fig1b}}
\end{figure}

In Figure \ref{fig1b} we show a fit for the three lines with $V_{\rm
in} = 7,000 \kms$ and $V_{\rm out} = 10,000 \kms$.  Given the
simplicity of the model, the fits are surprisingly good, especially
that of H$\alpha$.  There is a minor asymmetry in the H$\alpha$ line,
where the red side would be better fit with $V_{\rm in} = 6,000 \kms$
and $V_{\rm out} = 10,000 \kms$. There is some indication that the
[N~II] line has a somewhat smaller width than H$\alpha$, with $V_{\rm
in} \approx 6,000 \kms$. Because of the lower S/N compared to
H$\alpha$, however, this is hardly significant.

Part of deficit on the red side of Mg~II is caused by the interstellar
Mg~I $\wl 2852$ absorption line from the host galaxy of SN 1993J (M81)
and the Milky Way \citep[see][]{dB93}. This is, however, unlikely to
explain the entire asymmetry.  Additional interstellar Fe~II
absorption lines may possibly contribute, but these should not be
stronger than the Mg~I absorption, and therefore they only marginally
affect the line. It is also difficult to understand the red deficit as
a result of an intrinsic asymmetry of the Mg~II-emitting region, since
this should coincide with the H$\alpha$ line, which does not show such
a pronounced effect. In Section \ref{sec_dust} we discuss a different
explanation based on optical-depth effects in the emitting region.

Of the other UV lines, only N~IV] $\wl 1486$ is isolated enough for
blending not to be important. The noise, however, makes the profile of
this line uncertain. With this caveat, the line can be fit with the
same line profile as for H$\alpha$.

For the region below 2000~\AA, which is of most interest for the CNO
abundance analysis, the blending of the lines requires a detailed
spectral synthesis to derive accurate line fluxes. Because of the
higher S/N of the N~III]---C~III] region compared to the N~IV]---C~IV
region, we concentrate on the former.  Based on line identifications
in UV spectra of other supernovae and other photoionization-dominated
objects, we have added lines from all ions in the interval
1600--2000~\AA\ at the appropriate wavelengths and adjusted the
relative fluxes to provide a best fit. The lines included are He~II
$\wl 1640.4$, O~III] $\wll 1660.8, 1666.2 $, N~III] $\wll
1746.8$--1754.0, Ne~III] $\wl 1814.6$, Si~III] $\wll 1882.7, 1892.0$,
and C~III] $\wll 1908.7$. The relative intensities of the individual
multiplet components are calculated for $T_e = 15,000$~K and $n_e =
10^8 \cmc$, values typical of the conditions in the outer ejecta. From
models of the ejecta one also expects Al~III to have a substantial
fractional abundance. Because Al~III $\wll 1854.7, 1862.8$ are
resonance lines, they in general have high optical depths, causing a
scattering of lines blueward of Al~III (especially the N~III and
Ne~III] lines) to longer wavelengths. We calculated this scattering by
a Monte Carlo procedure, taking the doublet nature of the lines into
account. For all lines we assume a boxy line profile determined from
N~II] $\wll 2139.68, 2143.5$ (see above). The Al~III scattering is
assumed to take place in the same region.

\begin{figure}
\plotone{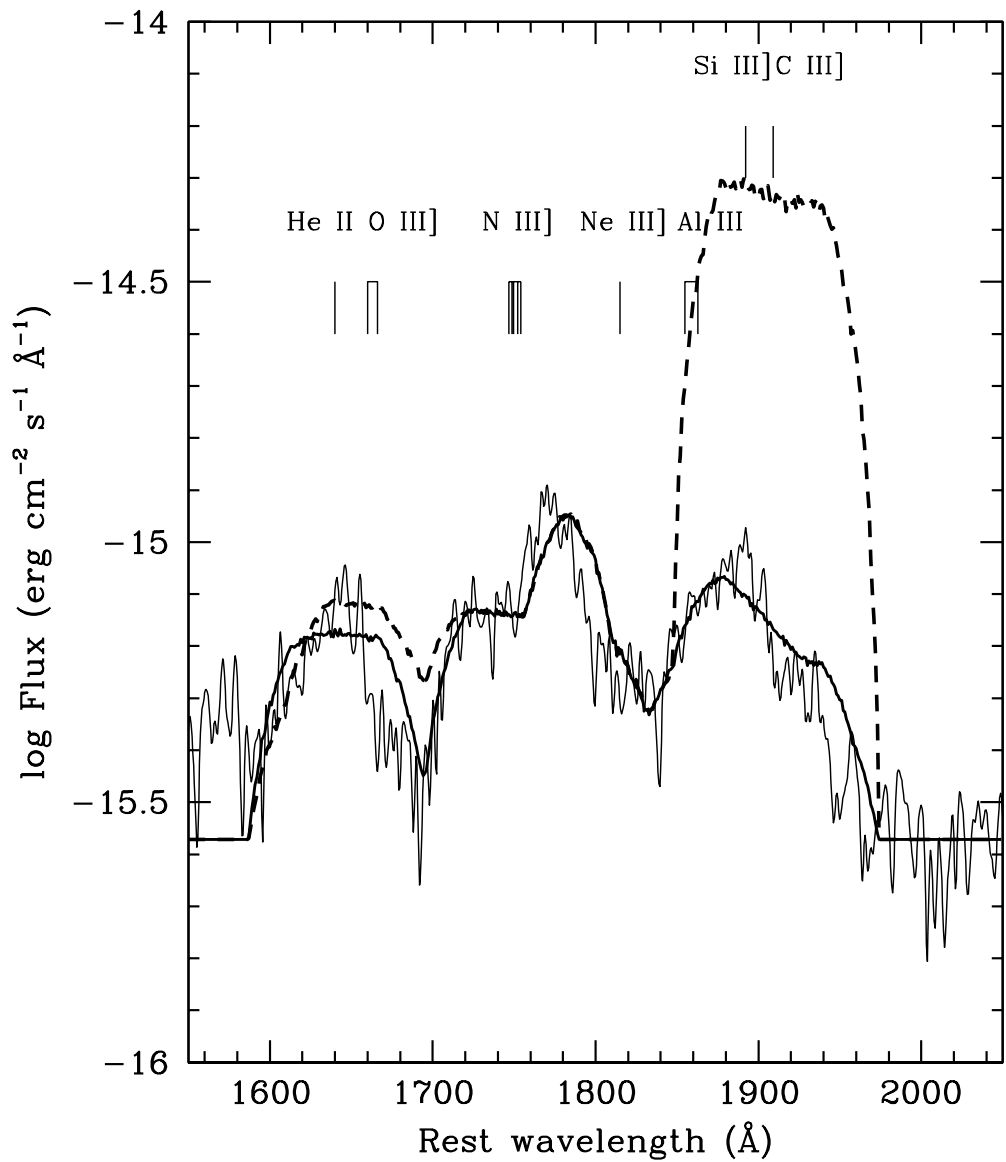}
\caption{Spectral synthesis model of the O~III], N~III], and C~III]
region of SN 1993J. The observed spectrum shown is the average of the
day 1063 (23 February 1996) and day 1399 (24 January 1997) spectra. 
The solid synthetic spectrum has N/C $=12.4$ and N/O $= 6$, while 
the dashed line shows the expected spectrum for solar CNO ratios, 
N/C $=0.25$ and N/O $=0.12$. }
\label{fig4}
\end{figure}

To increase the S/N in the abundance analysis we again use the
averaged day 1063 and day 1399 spectrum. The result of the best-fit
model is shown in Figure \ref{fig4}, and the individual line fluxes
relative to the total flux of the N~III] multiplet are given in Table
\ref{tab2a}.  Because of blending and resonance scattering, the
resulting spectrum has a complicated form. We note that most of the
feature at 1620--1680~\AA\ is due to He~II, rather than to O~III]. The
latter flux is uncertain, and we can only give a conservative upper
limit of $\sim1.0$ times the N~III] flux. The best fit in Figure
\ref{fig4} has an O III] flux only $\sim0.1$ times the N~III]
flux. The peak at $\sim1780$~\AA\ is caused by resonance scattering of
Ne~III] $\wl 1814.6$ by Al~III $\wll 1854.7, 1862.8$. The scattered
Ne~III] flux gives rise to some of the emission at $\sim1880$~\AA.
The uncertainties in the fluxes of the individual lines are difficult
to quantify, but we estimate the N~III] and the C~III] fluxes to be
accurate to $\pm 30\%$.

Finally, we comment on the N~IV] $\wl 1486$ and C~IV $\wll 1548.9,
1550.8$ lines (see Fig. \ref{fig00}). While the N~IV] line is clearly
seen, the C~IV doublet is unfortunately swamped by the continuum,
as well as affected by the interstellar absorption. We can
therefore only give a lower limit to the (N~IV] $\wl 1486$)/(C~IV
$\wll 1548.9, 1550.8$) ratio of $\sim1.5$. This limit is
conservative, because we have assumed a lower continuum for C~IV
than for N~IV]; it is likely that the true ratio is considerably
higher.

\subsection{SN 1998S}

\begin{figure}
\epsscale{0.85}
\plotone{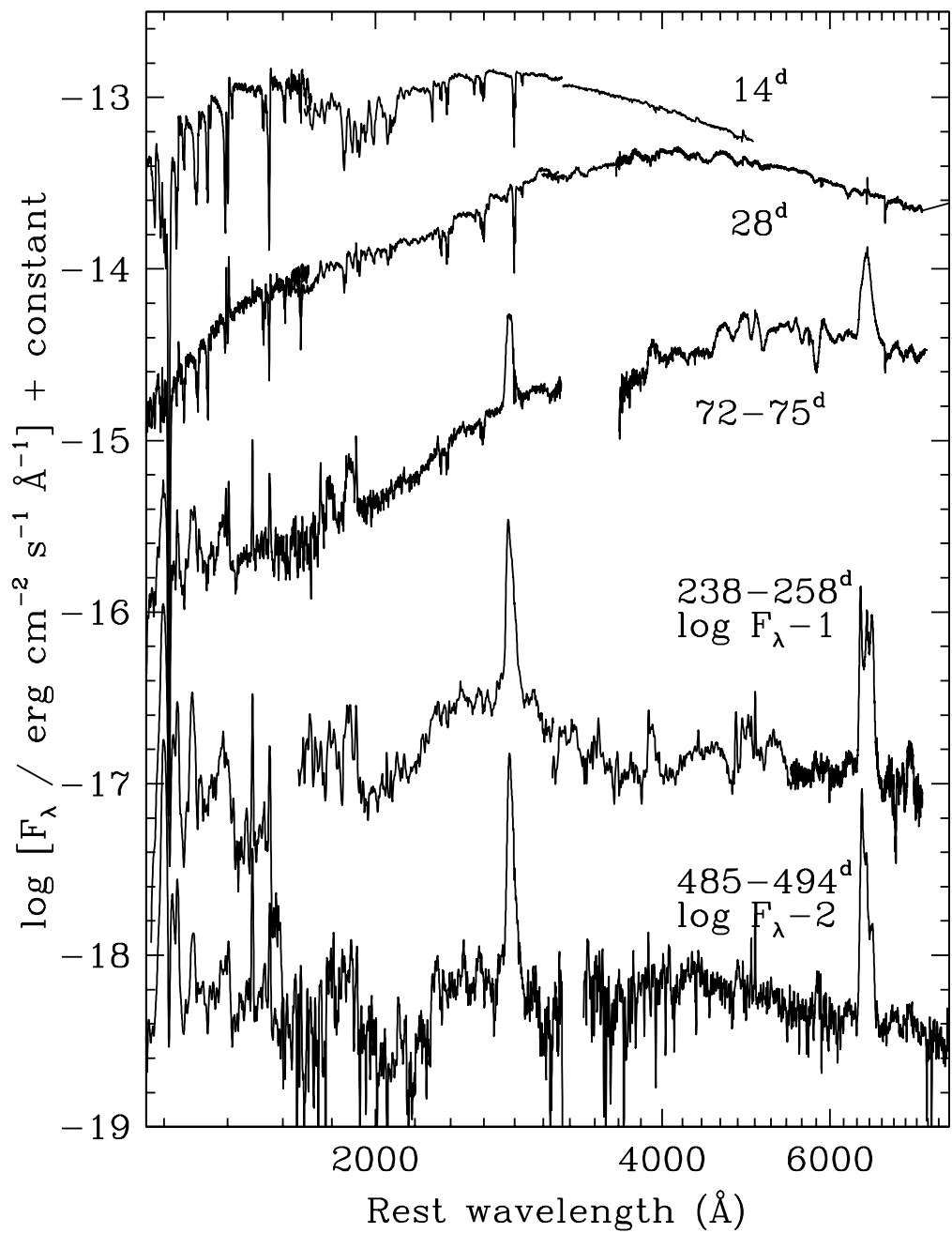}
\caption{Evolution of the spectrum of SN 1998S, with epoch (days after the
explosion) indicated. The lower two spectra are shifted downward by 1 and 2
dex, respectively, relative to the previous ones.
\label{fig0}}
\end{figure}

\begin{figure}
\plotone{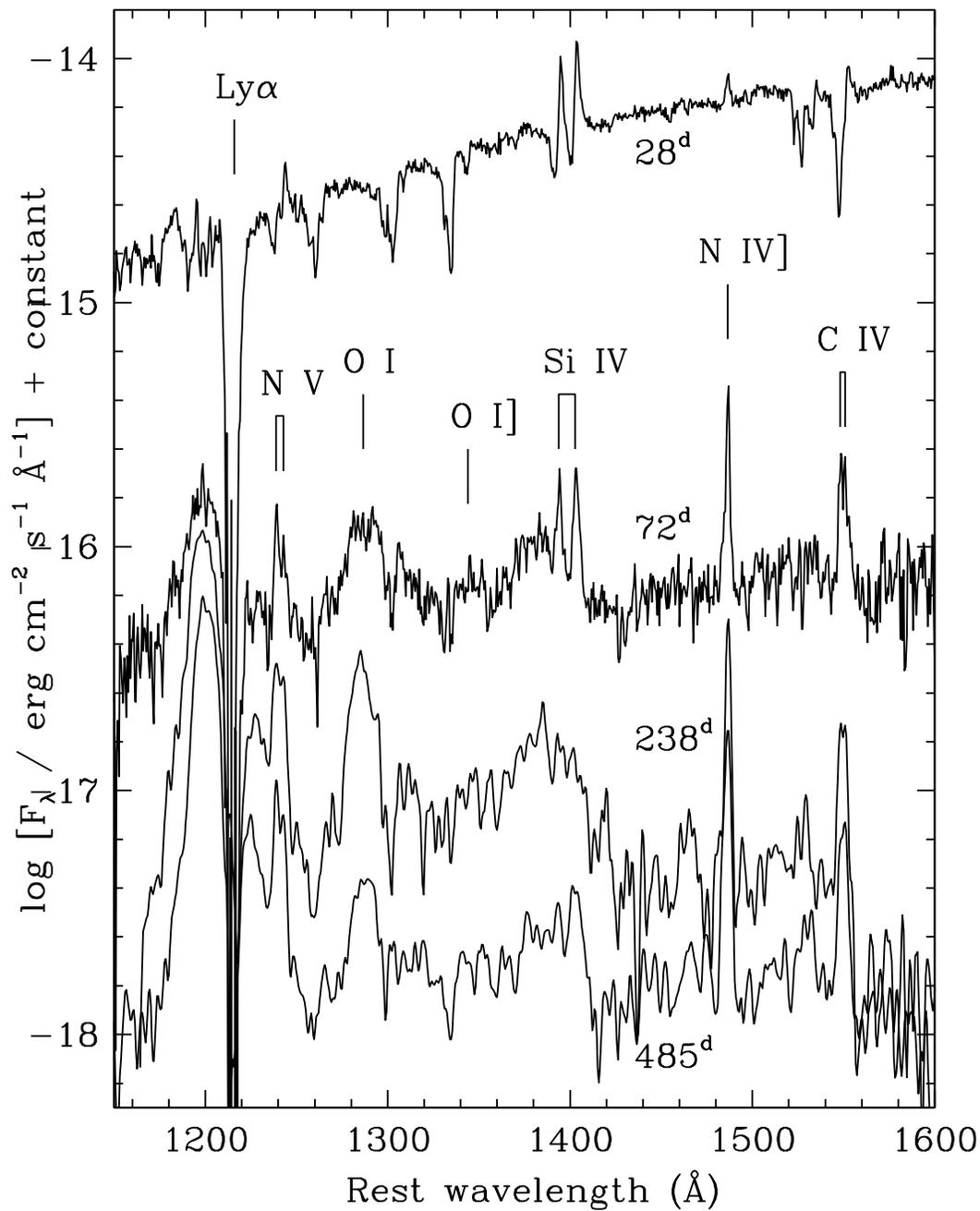}
\caption{Evolution of the far-UV spectrum of SN 1998S from 30 March
1998 to 30 June 1999.  Each spectrum is shifted down by 0.5
dex with respect to the previous one.
\label{fig1}}
\end{figure}

In Figure \ref{fig0} we show the full spectral evolution of SN 1998S
for the epochs where {\it HST} observations
exist, while in Figure \ref{fig1} the important far-UV region is shown
in more detail. Finally, we show in Figure \ref{figlateopt} the spectral
evolution in the optical range for the nebular phase. This
complements the observations in \citet{Leo00} by adding the very late
spectra at 643 and 2148 days. 

The spectrum during the first 1--2 months is mainly a continuum with
superimposed P-Cygni lines, as discussed by \cite{Len01}. By day 72
several strong emission lines have emerged. The lines can be divided
into narrow lines with full-width at half-maximum (FWHM) $\la 300
\kms$, and broad lines extending up to $\sim10,000\kms$ (half-width
near zero intensity; HWZI); they are likely to originate in the
circumstellar medium and in ejecta gas, respectively.

\subsubsection{Broad Lines}
\label{sec_broad}

The strongest of the broad lines are Ly$\alpha$, H$\alpha$, and Mg~II
$\wll 2795.5, 2802.7$.  In Figure \ref{fig1c} we compare their
profiles throughout the course of the {\it HST} observations. The flux
scale is linear and varies for the different lines and dates to
facilitate a comparison. During the first two epochs only the central
absorptions of Ly$\alpha$ and Mg~II are visible.  These features are
shown in detail in high-resolution spectra published by \cite{BRMB00},
where both a Galactic and a host-galaxy component are seen, in
addition to a $\sim350 \kms$ component from the circumstellar medium
of SN 1998S. Ly$\alpha$ is even more affected by interstellar
absorptions. Because of the large H I column density, these lines show
strong damping wings, extending $\sim10$~\AA\ from the line center.

On day 72 and later, both Ly$\alpha$ and Mg~II have developed a strong
blue emission component.  The shapes of the Ly$\alpha$ and Mg~II lines
are similar within the uncertainties, with the blue wing at least a
factor of 10 stronger than the red. The lines extend on day 72 to
$\sim8,000 \kms$ (HWZI). In the later spectra a marginal decrease in
this velocity can be seen.

We immediately note the difference between the H$\alpha$ line profile
on the one hand and Ly$\alpha$ and Mg~II on the other hand. While the
former line up to the 238--258 day spectrum shows a fairly strong
red component, this is missing for the latter. Further, in the day 
72--75 spectrum the H$\alpha$ line is peaked at $\sim -1,000 \kms$, and
extends to zero velocity.  The flux of both Ly$\alpha$ and Mg~II,
however, drops close to zero already at $\sim2,000 \kms$. As is seen
from the earlier spectra, one reason for this difference is likely to
be interstellar Ly$\alpha$ and Mg~II absorption from our Galaxy and
from the host galaxy.  Even considering this, the low fluxes in the
red wings of Ly$\alpha$ and Mg~II make them distinctly different from
H$\alpha$.

The evolution of the shape of the H$\alpha$ line has been discussed in
several papers \citep{Leo00,Ger00,Fas01,Poz04}. Our observations,
however, cover a longer time interval. In Figure
\ref{fig_ha_hb} we show the temporal evolution of the H$\alpha$ line
from day 75 to day 2148.  From being symmetric during the first
months, as seen in the day 75 spectrum, the line developed a clear
asymmetry by day 100, with the blue wing considerably fainter than the
red. By day 200 the line again changed character, now with a
considerably fainter red wing compared to the blue, as seen in the day
258 spectrum \cite[see also Fig. 4 in][]{Ger00}. This evolution
continued, and by day 653 the red wing was almost gone. In the very
late spectrum on day 2148 the blue peak had almost disappeared, and
instead the central peak was strong. The
red part of the line is still much weaker than the blue. The absolute
level, however, is uncertain due to the difficulty of an accurate
background subtraction. 

A central peak is seen in the spectra at several epochs. It is,
however, not clear that this peak is of supernova origin. The flux of
this peak relative to the rest of the line is sensitive to the
subtraction of the background emission from the galaxy, and therefore
to the atmospheric seeing. In particular, the reality at the last
epochs is questionable. In these the background [N II] $\wll 6548.0,
6583.5$ lines at $-686 \kms$ and $960\kms$, respectively, can clearly
be seen, which shows that there is contamination from the
background. The fact that a peak is not seen in the Mg II line,
observed with {\it HST}, further strengthens this conclusion.

From Figure \ref{fig_ha_hb} we note the clear decrease in the velocity
of the blue peak from $\sim$4,300--4,700 $\kms$ on days 108--258 to $\sim
3,900 \kms$ on days 494 and 653. The maximum blue velocity decreases
marginally from $\sim7,300 \kms$, to $\sim6,800 \kms$.  In addition to
the dramatic change in the line profile, the large Balmer decrement is
of special interest, as noted earlier by \cite{Leo00}, who argue that
this strongly implies an origin in gas of very high density.

By day 72 the Mg~II line already showed a strong red-blue asymmetry,
which persisted throughout the duration of the remaining observations,
but there is no indication of a red peak in this line. The Ly$\alpha$
line, however, shows an indication of a red peak in the spectra after
day 72.  \cite{Ger00} propose that the fading of the red peak relative
to the blue can be explained as a result of dust formation. The fact
that the red wings of Mg~II and Ly$\alpha$ are suppressed earlier than
H$\alpha$ is consistent with this interpretation, as is the dust
signature seen in the infrared \citep{Ger02}. For a standard
extinction curve, the optical depths at Mg~II and Ly$\alpha$ are
related to that at H$\alpha$ by $\tau({\rm Mg~II}) = 2.4\, \tau({\rm
H} \alpha)$ and $\tau({\rm Ly}\alpha) = 4.3\, \tau({\rm H}\alpha)$.

Other broad lines at $\sim1288$, 1344, 1380, 1790, and 1880~\AA\ are
less obvious to identify. We note, however, the similar profiles of
these lines to that of the Mg~II line. Based on the velocity shift of
the blue peak, we find that a consistent set of line profiles can be
obtained if we identify these lines as O~I $\wll 1302.2$--1306.0, O~I
$\wll 1355.6$--1358.5, Si~IV $\wll 1393.8, 1402.8$, [Ne~III] $\wl
1814.6$, and C~III] $\wll 1906.7$--1908.7, respectively. The presence
of broad O~I lines in the UV is not surprising in view of the broad O
I $\wll 7771.9$--7775.4 features identified by \cite{Fas01}; both sets
of lines probably arise as a result of recombination. We return to the
implications of the presence of these lines in \S \ref{sec_origin}.

In Figure \ref{fig3b} we show the day 72--75 optical and UV line
profiles on a common velocity scale. There is some indication for a
broad component of C~IV $\wll 1548.9, 1550.8$, but this is
marginal. For N~V $\wll 1238.8, 1242.8$, N~IV] $\wl 1486$, and N~III]
$\wll 1746.8$--1754.0 there is no evidence for any broad components
above the noise. From the figure we note that there is marginal
evidence for a somewhat smaller extension of the blue wing of
H$\alpha$ (to $\sim7,000 \kms$) compared to Mg~II. The Ly$\alpha$,
C~III], Si~IV, and O~I lines have velocities consistent with the Mg~II
line. The [O~I] $\wll 6300.3, 6363.8$ lines seem to be considerably
narrower, but their shape is affected by neighboring lines, as well as
by the large separation of the doublet components.

\begin{figure}
\epsscale{0.85}
\plotone{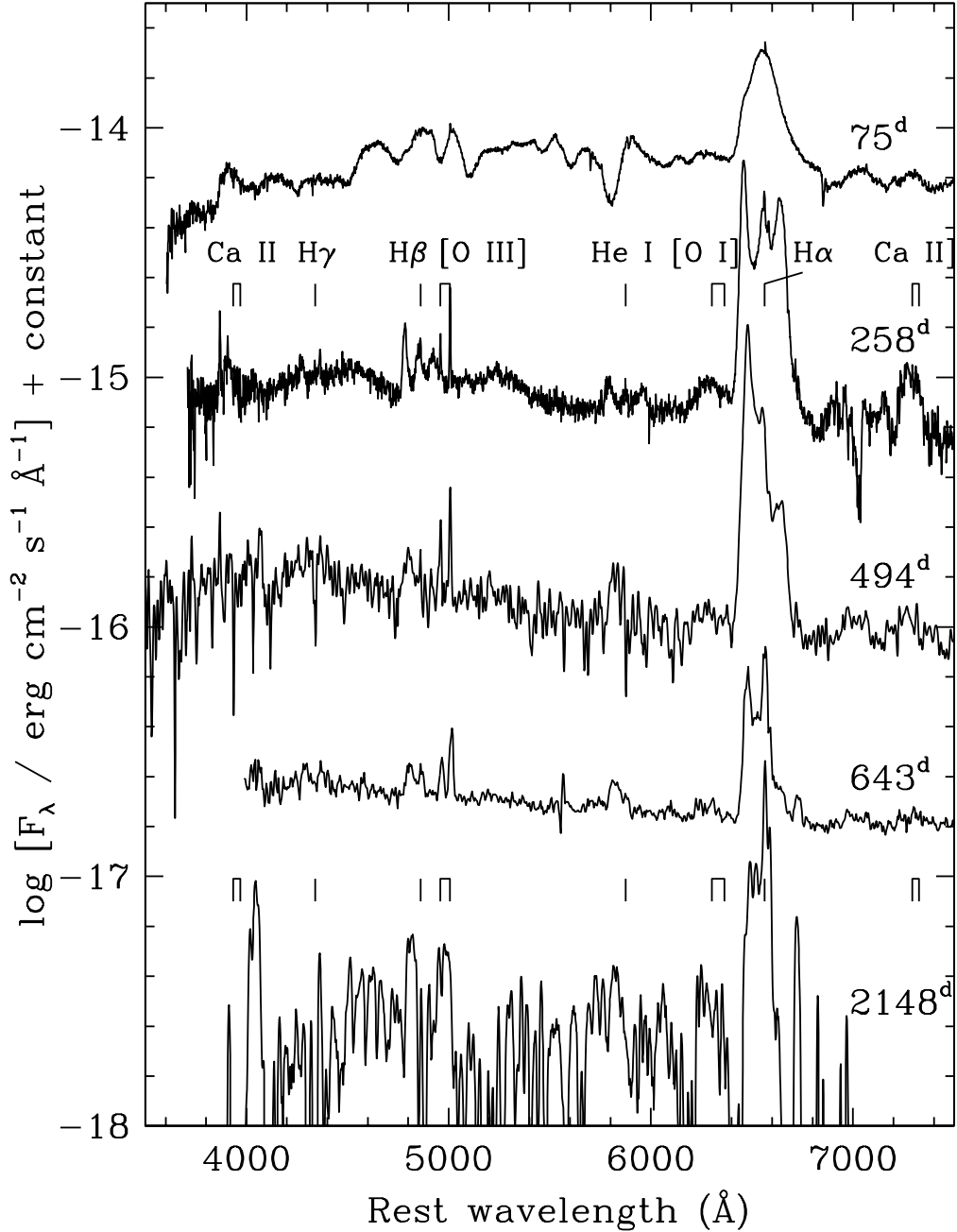}
\caption{Evolution of the optical spectrum of SN 1998S in the nebular
phase, where rest wavelengths of the most important lines are
marked. The spectra are not calibrated on an absolute scale, but the 
relative flux calibration within each spectrum is generally reliable.
The spectrum at 643 days was taken under bad seeing and is contaminated by
background H II regions, decreasing the contrast of the lines, as well
as adding narrow components at the rest wavelength of especially \Ha\
and [O III] $\wll 4959, 5007$. Also the day
2148 spectrum is contaminated by the galaxy background, and we have
only been able to make an approximate subtraction of this.
\label{figlateopt}}
\end{figure}

\begin{figure}
\plotone{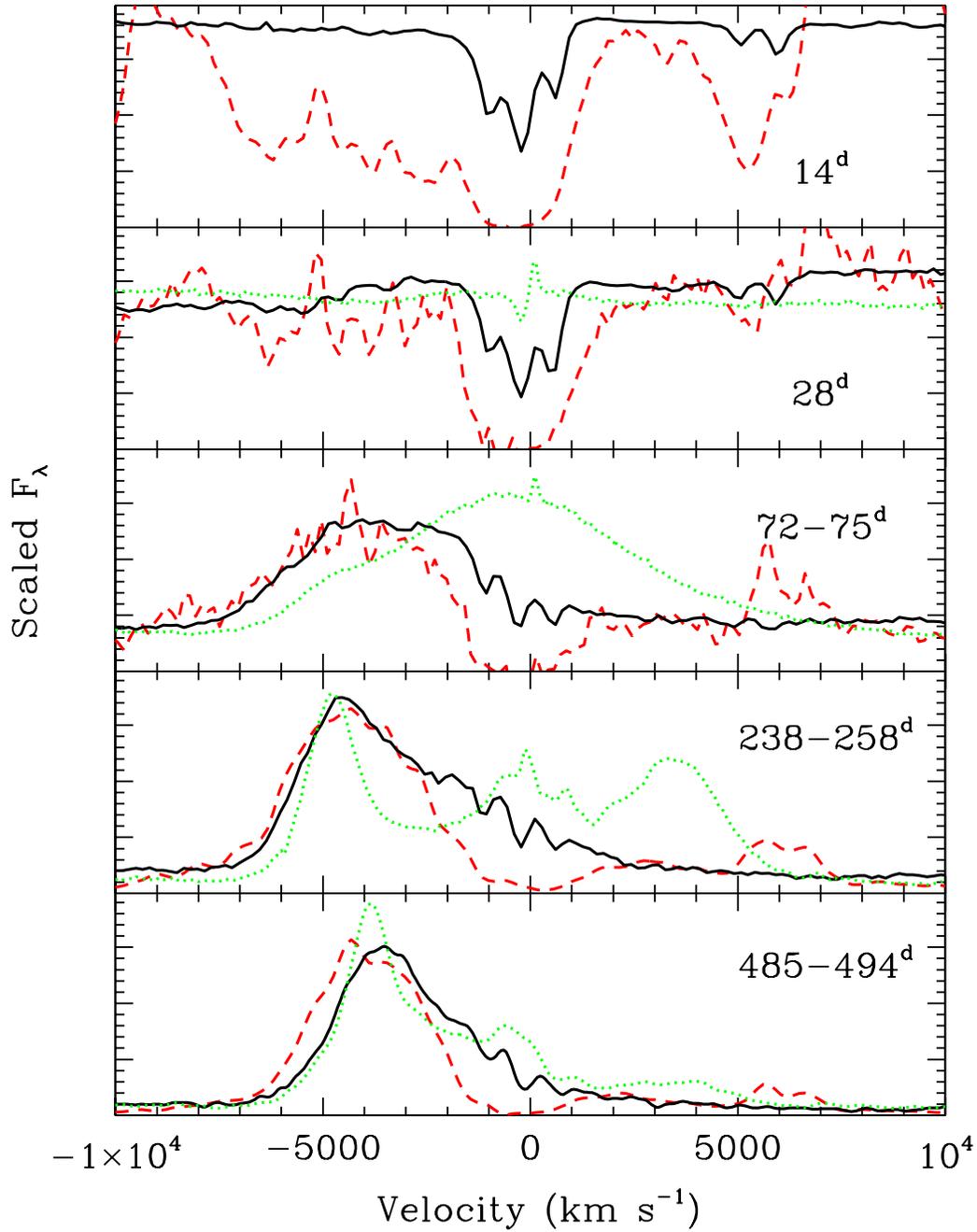}
\caption{Evolution of the Mg~II (solid lines), Ly$\alpha$ (dashed lines), and
H$\alpha$ (dotted lines) profiles in SN 1998S from 16 March 1998 (day 14) to 30
June 1999 (day 485).  The H$\alpha$ region was not observed at the first epoch
(day 14).
\label{fig1c}}
\end{figure}

\begin{figure}
\plotone{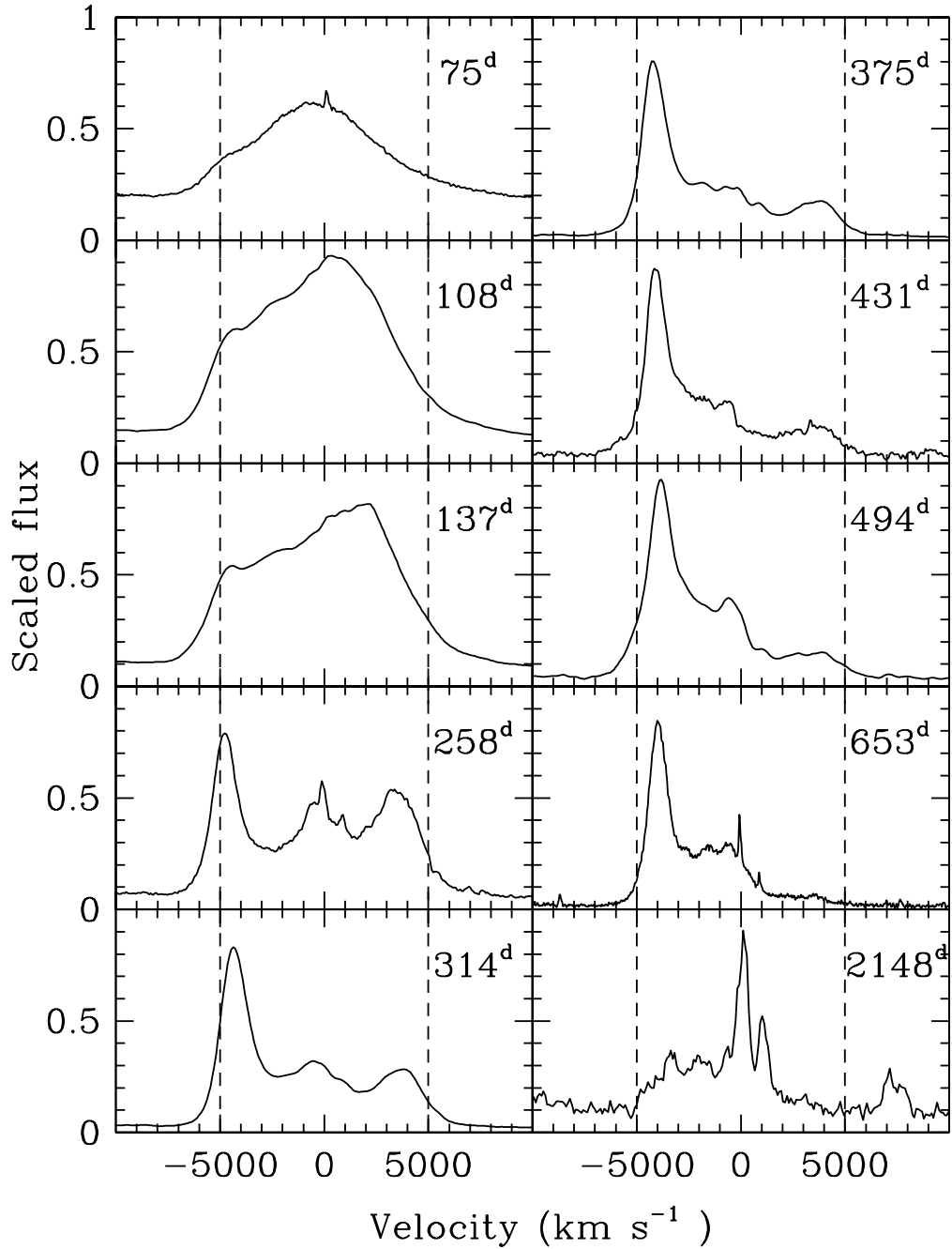}
\caption{Evolution of the H$\alpha$ line in SN 1998S. The linear flux
scale differs among the different dates. The dashed lines mark $\pm
5,000 \kms$. Note the progressive decrease in the velocity of the blue
peak from $\sim 4,500 \kms$ to $\sim 3,900 \kms$.
\label{fig_ha_hb}}
\end{figure}

\begin{figure}
\plotone{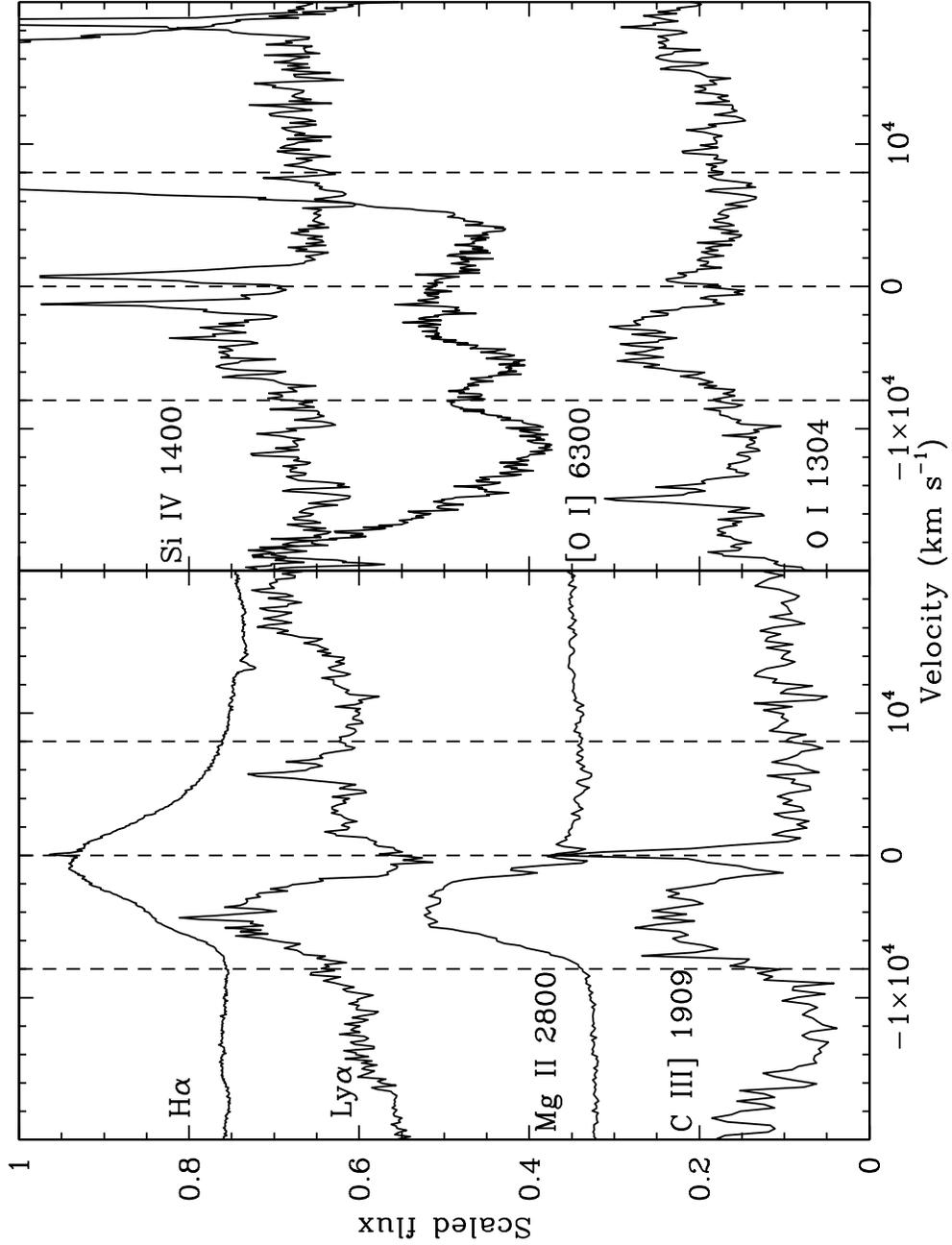}
\caption{Comparison of the SN 1998S 
line profiles of H$\alpha$, Ly$\alpha$, Mg~II,
C~III], Si~IV, [O~I], and O~I $\wll 1302.2$--1306.0 on day 72. To show the
profiles more clearly the flux scale differs among the lines. The dashed
vertical lines show a velocity of 0 and $\pm 8,000 \kms$ from the rest
wavelengths.
\label{fig3b}}
\end{figure}

The [O~III] $\wll 4958.9, 5006.8$ region is similarly complicated by
the doublet nature, as well as by the neighboring \Hb\ line. In the
spectra from 75--137 days, there is a clear feature at approximately
the position of the [O~III] lines (Figure \ref{figlateopt}). The peak
is $\sim2000 \kms$ redward of the rest wavelength, but this is also
approximately the case for \Ha. To test this more quantitatively we
have used the \Ha\ profile to create a synthetic spectrum, including
\Hb\ and [O~III] $\wll 4958.9, 5006.8$, adjusting the line strengths
so as to give a best fit to the 4700--5200~\AA\ region. The result of
this, however, gives a peak that is too blue, at $\sim5040$~\AA. In
addition, the [O~III] $\wll 4958.9$ component gives a too shallow red
wing of the \Hb\ line. We therefore conclude that the feature at
$\sim5040$ \AA\ in the early spectra is unlikely to be due to [O~III],
but is probably an Fe~II line with $\wl \approx 5015$~\AA.  In the
spectra later than $\sim300$ days there is a faint line at the [O~III]
wavelength, best seen in the last spectrum at 2148 days, which is most
likely from [O~III].  The line profile of the [O~III] line is, within
the errors, similar to that of \Ha, with a heavily absorbed red
component.

Other lines in the optical spectrum in Figure \ref{figlateopt}
are discussed by \cite{Ger00}, \cite{Leo00}, \cite{Fas01}, and
\cite{Poz04}. We only remark that
in our last two spectra on days 643--653 and 2148 we see the same lines as
in the earler spectra, in particular H$\beta$, He I $5876$, [O I]
$\wll6300, 6364$, and [O III] $4959, 5007$.

\subsubsection{Narrow Lines}
\label{sec_nl}

Because of the relative lack of blending, the identification and flux
measurements of the narrow circumstellar lines are considerably
simplified compared to the broad lines.  This is particularly true
when we compare with the flux measurements of the far-UV lines of SN
1993J, where we had to use a complicated deblending procedure.

In Table \ref{tab2} we give the fluxes of the narrow lines below
2300~\AA.  Because of the strong continuum and the influence of
P-Cygni absorptions of several lines, the day 28 fluxes are
uncertain. The noise level of the days 238 and 485 spectra are such
that the fluxes are also very uncertain for many of the lines. In the
subsequent abundance analysis we therefore restrict ourselves to the
day 72 spectrum, which has a suitable combination of high S/N and
well-defined emission lines (see Fig. \ref{fig1}).

\begin{figure}
\epsscale{0.85}
\plotone{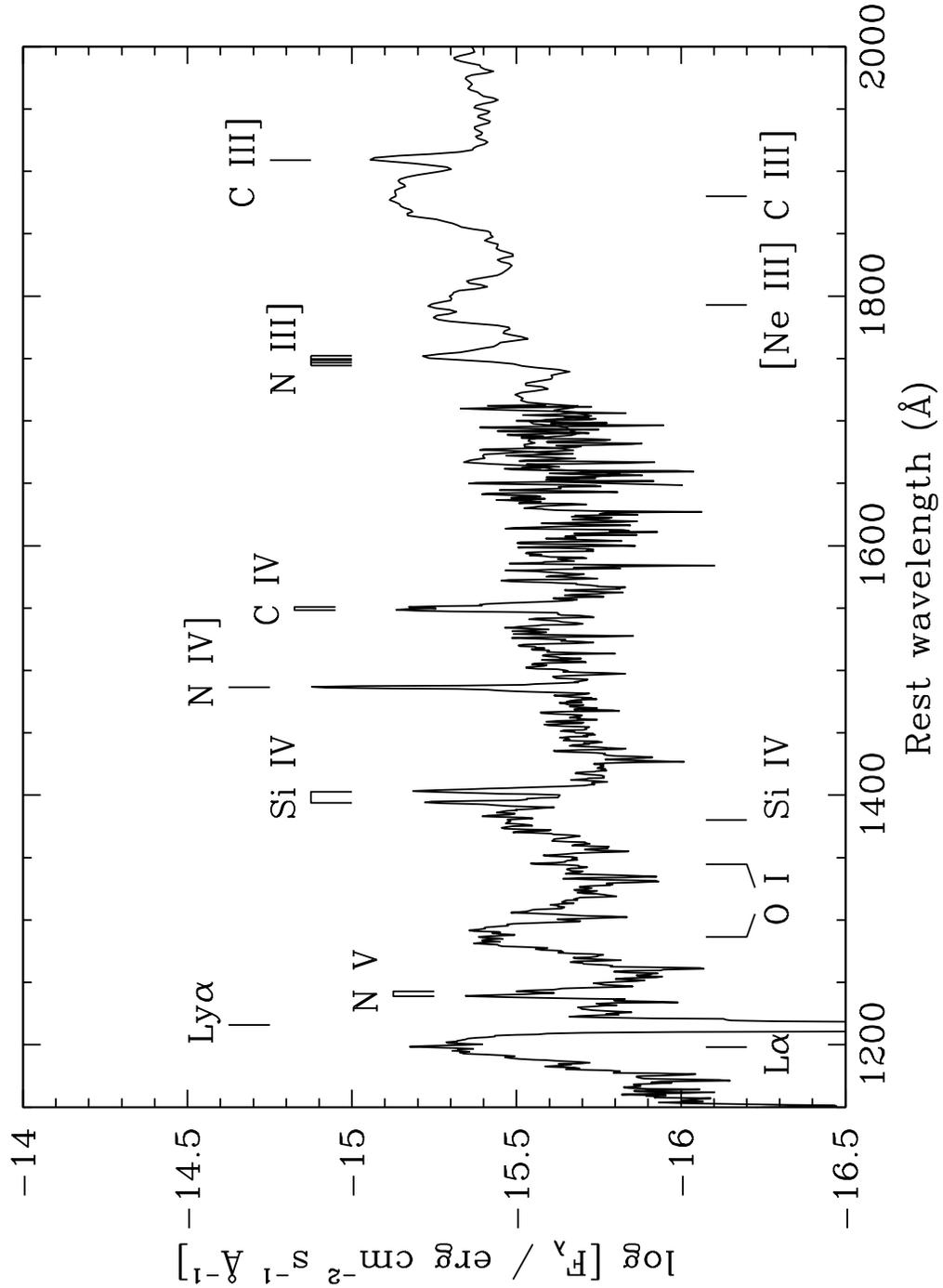}
\caption{The far-UV spectrum of SN 1998S on 13 May 1998 (day 72).
Identifications of the narrow lines are given in the upper part of the figure
(except for \La). In the lower part identifications of the broad lines are
given. The lines here show the observed center of the feature, not the rest
wavelengths.
\label{fig2}}
\end{figure}

In Figure \ref{fig2} we show the day 72 spectrum shortward of
2000~\AA\ in greater detail, together with line identifications.
Based on their observed rest wavelengths in the day 72 spectrum,
1394.0~\AA\ and 1403.1~\AA, we identify these lines as coming from
Si~IV rather than O~IV. The FWHM of N~IV] $\wl 1486$ is consistent
with being unresolved, noting that the resolution is $\sim250 \kms$ at this
wavelength.  High-resolution optical and UV observations during the
first month showed two low-velocity components
\citep{Fas01,BRMB00}. While [O~III] $\wl 5006.8$, for example, had a
FWHM of 50--60 $\kms$, the intermediate-velocity H$\alpha$ absorption
component extended to $\sim350 \kms$. Unfortunately, the limited
resolution of our spectra prevents us from relating the narrow
emission lines to one or the other of these components. The C~IV $\wll
1548.9, 1550.8$ doublet is clearly resolved in the day 72
spectrum. Also, the Si~IV $\wll 1393.8, 1402.8$ and the N~V $\wll
1238.8, 1242.8$ doublets are resolved in the same spectrum. The
intensity ratio (N~V $\wl 1239$)/(N~V $\wl 1243$) is $\sim1.5$.  In
the region longward of 2000~\AA, narrow lines of [N~II] $\wll 2139.7,
2143.5$ are clearly present, as well as C~II] $\wll 2322.7$--2328.8.

For the following discussion we note that the uncertainties in the
fluxes in the day 72 spectrum are 7--9\% for all lines, except for the
N~III] lines ($\sim20$\%). For the other dates the uncertainties are
considerably larger. These values only include the noise, estimated in
the neighborhood of each line. In addition, there may be systematic
errors originating from the assumed level of the continuum, as well as
from line blending; in contrast to SN 1993J, however, these errors are
likely to be small because of the well-defined line profiles.

As for SN 1993J, the high noise level in the 1600--1700~\AA\ region
prevents us from giving a reliable O~III] $\wl 1664$ flux. As an upper
limit to the flux we find $\sim1.5\times 10^{-15} \ergs {\rm
~cm}^{-2}$.

\section{DISCUSSION}

\subsection{Origin of the Emission Lines}
\label{sec_origin}

As has been discussed elsewhere \citep[e.g.,][]{CF94, CF03}, the
origin of the broad lines seen in SN 1993J and SN 1998S can be
explained as a result of photoionization by X-rays and UV emission
from the radiative reverse shock propagating into the supernova
ejecta. The reprocessed radiation from the low-density, unshocked
ejecta emerges mainly as UV lines of highly ionized species like
C~III-IV, N~III-V, and O~III-VI. Because of cooling, a dense shell is
formed behind the reverse shock --- that is, between the reverse shock
and the contact discontinuity. The high density in this region causes
the reprocessed radiation to emerge mostly as emission from neutral or
singly ionized ions, such as H~I lines, Mg~II, and Fe~II.

While the box-like line profiles in SN 1993J (Fig. \ref{fig1b}) are
consistent with those coming from a narrow, spherical shell, the
geometry of the line-emitting region in SN 1998S is less
clear. \cite{Ger00} interpret the double-peaked H$\alpha$ line profile
as a result of the interaction of the ejecta with a circumstellar ring
\citep[see also][]{CD94}. The centrally peaked profile is explained by
a separate component --- a spherical distribution of shocked
circumstellar clouds. Similar ideas are proposed by \citet{Leo00}, who
relate this picture to a binary scenario, reminiscent of that
discussed for SN 1987A.

Although there is no quantitative discussion of this hypothesis, some
features in this scenario may be understood from the general picture
in \cite{CF94}. The broad wings would then represent the emission from
the ejecta, while the narrow component is from the shock propagating
into the ring, and possibly also the reverse shock, in case this is
radiative. Most of the UV and optical line emission is then produced
as a result of photoionization of the ejecta, as well as the ring and
the cool postshock gas. The ingoing fraction will mainly be absorbed
by ejecta gas close to the reverse shock, explaining the ring-like
emission also from the ejecta component. The density distribution of
the ejecta is likely to be more uniform with respect to polar
angle, possibly resembling the non-spherical models of
\cite{BLC96}. We return to this scenario in the next section.

In SN 1995N \citep{FC02}, there was a clear distinction in the line
widths of the hydrogen and Mg~II lines compared with those of the
high-ionization UV metal lines. While the former had smoothly
declining line profiles with velocities (HWZI) up to $\sim10,000
\kms$, the latter were box-shaped with HWZI $\approx 5,000 \kms$. No
such distinction is obvious either for SN 1993J or SN 1998S.

The strong O~I recombination lines present in SN 1998S resemble those
seen in SN 1995N. In \cite{FC02}, it was argued that in order to
explain their strengths relative to the hydrogen lines, they had to
originate in processed gas with a high abundance of oxygen. We believe
the same is true for SN 1998S as well, which indicates that some of
the gas giving rise to the broad metal lines comes from processed
gas. SN 1993J also has strong metal lines in the UV, as well as in the
optical, which grow stronger with time relative to H$\alpha$
\citep{MFb00}. Hence, there cannot be a large mass of hydrogen present
between the reverse shock and the oxygen-rich core. It is likely that
SN 1993J, SN 1995N, and SN 1998S had lost most of their hydrogen-rich
gas, and the helium and oxygen core is therefore efficiently
illuminated by the X-rays from the reverse shock.

The origin of the relatively narrow lines seen in SN 1998S, and in other
SNe~IIn, is less clear. Given their small velocity width, they must arise in
circumstellar gas. This can either be unshocked gas in front of the shock, or
dense regions shocked by the blast wave.

\subsection{Asymmetric \Ha\ Emission and Dust Formation in SN 1998S}
\label{sec_dust}

As mentioned in \S \ref{sec_intr}, there are several indicators of
dust present in SN 1998S, both from the line profiles and from the
strong infrared excess \citep{Ger02,Poz04}. In addition, the double-peaked
line profiles provide a very interesting clue to the nature of the
line-emitting region. \cite{Ger00}, in particular, attribute this to
emission by a torus-like region. In this section we discuss the dust
and the geometry of the line-emitting region.

The location and nature of the dust is not clear. An obvious candidate
is dust formed in the heavy-element enriched ejecta. If most of the
line emission comes from a shell close to the reverse shock, the dust
in the core will primarily absorb the high-velocity emission from the
receding ejecta, in agreement with the observations. This is the
scenario favored by \cite{Ger00} for the asymmetric line
profiles. \cite{Ger02}, however, find that the same dust cannot
explain the excess emission seen in the infrared. Instead, they
attribute this to an echo from pre-existing dust.

Another interesting possibility is that the dust absorption, as well
as the emission, takes place in the cool dense shell (hereafter CDS)
in the shock region \citep{DCH03}. This is a favorable place for dust
formation because of its high density and low temperature
\citep{F84a,CF94}. Whether it is cool enough, given the X-ray heating
from both shocks, remains to be demonstrated quantitatively. The fact
that the gas swept up by the shocks may be enriched by heavy elements
may be important in this context. Indications for metal enrichment in
SN 1998S are also presented from X-ray observations by
\cite{Poo02}. The temperature of the emission should be close to the
condensation temperature, which agrees with the \cite{Ger02}
temperature estimate of the dust emission, 700--1400 K. Recently,
\cite{Poz04} find that the luminosity and temperature of the IR
emission gives a blackbody radius corresponding to a velocity of $\sim
4,000 \kms$ at 330--400 days. This is comparable to the velocity of
the blue and red peaks of the \Ha \ line at this time. Later, the
blackbody velocity of the dust receeds to $\sim 1,000 \kms$, indicating
a thinning of the dust.

A problem with this idea is that the dust is likely to condense behind
the zone relative to the reverse shock where H$\alpha$, Mg II, and
other low-ionization lines are formed. If the CDS is formed behind the
reverse shock (in the Lagrangian sense), this means that both the
front and back side line emission will be absorbed by the dust,
contrary to the observations. A solution to this may come from the
fact that hydrodynamical mixing may give rise to a clumpy CDS
\citep{CB95}. This could lead to a covering factor of the dust less
than unity, and allow some of the \Ha\ to penetrate.

An alternative, related possibility, discussed for SN 1993J by
\cite{FLC96}, is that the forward shock is radiative. In this case the
line emission from the receeding part of the shell (red wing) will be
absorbed by the dust shell, while the line emission from the front
(blue wing) can emerge freely. Although dismissed for SN 1993J, a high
circumstellar density, as may be the case for SN 1998S, may make this
a real possibility. Alternatively, there may be a combination of
ejecta dust absorption and emission from dust created in the CDS. The
latter then has to be either clumpy or optically thin, in order to not
block the line emission from the shell.

Finally, we note that emission from perhaps 2--4~$\Msun$ of cold dust
from Cas~A has recently been claimed by \cite{Dun03}. This estimate of
the amount of dust, however, has been questioned by \cite{Dwek04}, who
finds that a more reasonable mass of only $10^{-4}-10^{-3} \Msun$ is
needed, if the dust is in the form of conducting, metallic
needles. This illustrates the sensitivity of the mass estimate to the
nature of the dust. But the most interesting result of \cite{Dun03} is
that the dust emission in Cas~A is concentrated in the region between the
reverse and forward shocks. This is consistent with the scenario above
where the dust is formed in the CDS during the first years after the
explosion.

To study the different scenarios for the dust absorption above in more
detail, as well as the effects of an asymmetric \Ha\ emission, we have
calculated several models for the line profiles. In particular, we
have considered the different possibilities discussed above for the
location of the dust-absorbing region.

To explain the double-peaked nature of the H$\alpha$ line, we assume
in this section that the emission originates in a torus-like region,
similar to that proposed by \cite{Ger00}. Some support for this also
comes from the polarization data for SN 1998S presented by
\cite{Leo00}. This region is characterized by the inner and outer
velocities ($V_{\rm in}$ and $V_{\rm out}$, respectively), the angular
thickness from the equatorial plane ($\theta$), and the inclination of
the torus relative to the line of sight ($i$, where $i = 90^\circ$ is
an edge-on torus). We assume the emissivity within this region to be
constant in both the radial and polar directions. The line is assumed
to be optically thin. 

To model the dust absorption we consider three different locations,
either inside the ejecta, behind the reverse shock, or behind the
circumstellar shock. We assume the dust to be opaque, but vary the
total areal covering factor, $f_{\rm c}$.

\begin{figure}
\epsscale{0.85}
\plotone{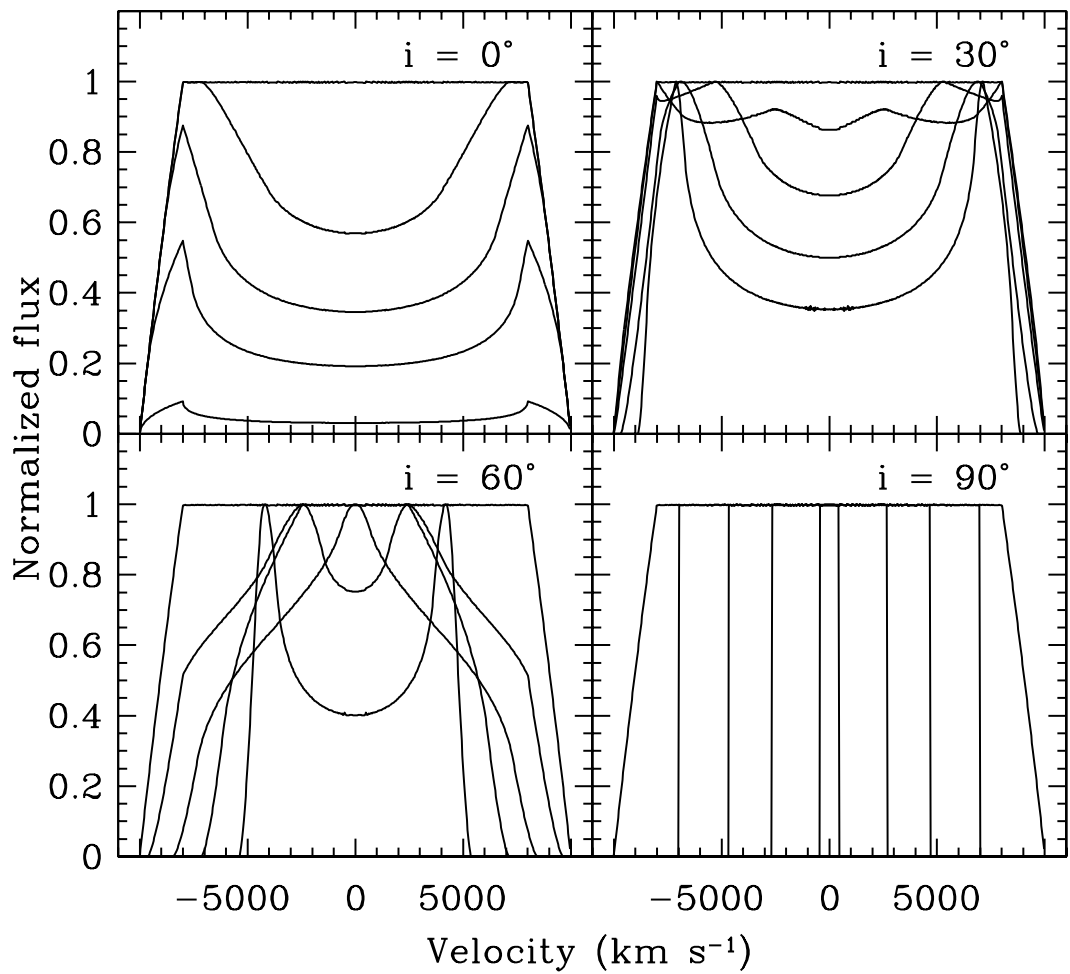}
\caption{Line profiles for a uniformly emitting torus for four different values
of the inclination, $i$, and for different values of the angular thickness of
the torus, $\theta=5\degr, 30\degr, 50\degr, 70\degr$, and $90 \degr$. In the
upper two panels, the flux at $v = 0$ is progressively higher as $\theta$
increases. In the lower two panels, the wings of the line go to progressively
higher velocities as $\theta$ increases. No dust absorption is
included.
\label{fig_lprof}}
\end{figure}

In Figure \ref{fig_lprof} we first show line profiles without
absorption, but at four different inclination angles ($i =0\degr,
30\degr, 60\degr$, and $90 \degr$) and for several values of the
angular thickness from the equatorial plane of the torus. We see here
that widely different line profiles may result, depending on the
inclination and angular thickness of the torus. In particular, both
double-peaked profiles and centrally peaked profiles may arise --- the
former mainly at low inclinations, while the latter occur for high
inclinations and high angular extent of the torus.

By varying the angular thickness of the torus we can create a sequence
of line profiles, which approximately resembles the observed evolution
of the \Ha\ line. In principle, one could therefore think of an
evolutionary time sequence from a large angular thickness to a smaller
one. This could, for example, result from the interaction of the
supernova with a flattened circumstellar environment, of which a thick
disk is the most extreme example. As the supernova expands, the
angular extent of the torus would then decrease, and the above evolution in
$\theta$ would result. We believe this to be the most promising of the
torus models.

To this emission we can now add dust absorption according to the
various possibilities discussed above. The cases with dust absorption
outside of the \Ha-emitting region all have the problem that the red
wing is considerably stronger than the blue wing. This conclusion is
independent of the inclination, the angular thickness, and the
covering fraction of the dust. Because this is in contradiction to the
observations, we do not further consider these cases.

\begin{figure}
\epsscale{0.80}
\plotone{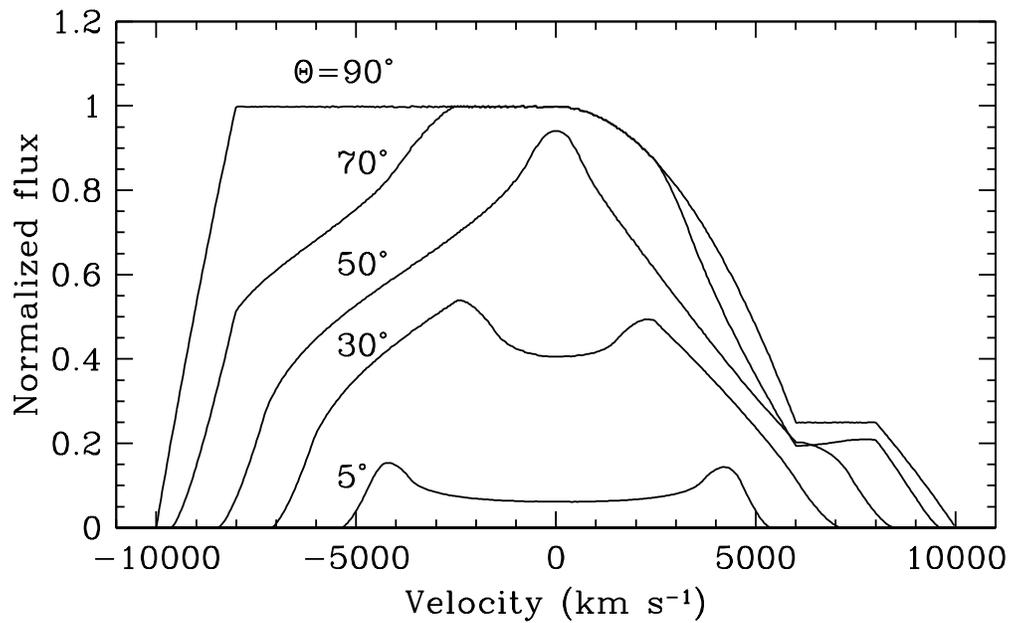}
\caption{Line profiles, including dust absorption, for different
values of the angular thickness of the torus for $i=60\degr$, $V_{\rm
in}=8,000 \kms$, and $V_{\rm out}=10,000 \kms$. The dust has been
assumed to be inside the \Ha-emitting torus with a covering factor
$f_{\rm c}=0.5$. Note the transition from a centrally peaked profile
to a double-peaked profile as the thickness of the torus decreases.
\label{fig_lprof_dust}}
\end{figure}

If we instead assume that the dust absorption occurs inside of the
\Ha-emitting region, as would be the case for a radiative forward
shock, the line profiles are more similar to those observed.  As an
example, we show in Figure \ref{fig_lprof_dust} a model with $V_{\rm
in}=8,000 \kms, V_{\rm out}=10,000 \kms$, $i=60 \degr$, $f_{\rm
c}=0.5$, and different values of the angular extent $5\degr< \theta<
90\degr$. The line profile with $\theta \approx 60\degr$ has an
approximate resemblance to the \Ha\ profiles on days 75--137 in Figure
\ref{fig_ha_hb}, while the profiles at the later epochs resemble those
of smaller $\theta$.

A major problem with this picture, however, is that the extent in
velocity of the blue wing is expected to decrease by a factor of about
two with time during this evolution, as shown in Figure
\ref{fig_lprof_dust}. The reason is simply that the velocity along the
line of sight decreases as the thickness decreases, and is difficult
to avoid. Also, the extent of the flat, weak red wing is difficult to
reproduce in this model.

Finally, we consider the case with absorption from dust formed in the
ejecta. A total obscuration by this dust is not compatible with the
presence of a weak, but finite red wing of the lines. A partial dust
obscuration, like the one in SN 1987A \citep{LDGB91} is, however, both
more realistic and compatible with the observations. To explain the
temporal evolution from a centrally peaked profile to a double-peaked
profile, we must refer to a situation similar to that in Figure
\ref{fig_lprof_dust}. Although the red part of the line may be better
reproduced by this configuration, it has the same problem with a too
rapidly decreasing line width as the model above. In addition, it is
difficult to reproduce the central peak without an additional
component, like the one invoked by \cite{Ger00}. The reality of this
peak is, however, not clear. 

In summary, we find it difficult to reproduce the observed evolution
of the line profiles with an emitting region in the form of a
torus. This is independent of the dust absorption, although more
complicated models cannot definitely be ruled out.

\subsection{Line Formation in the Cool Dense Shell}
\label{sec_lcds}

While the line profiles of the broad lines in SN 1998S are very clear
cases of double-peaked profiles, this also applies to the \La, Mg~II,
and \Ha\ lines in SN 1993J (see Figure \ref{fig1a}). In the case of SN
1993J there are no indications of major non-spherical structures. On the
contrary, VLBI imaging shows a basically spherical shell
\citep{Marc97,BBR01}, which excludes a flattened structure unless we
are very close to the axis of symmetry. Therefore, the torus
model for the peaks in the line profile is in the case of SN
1993J unlikely also for this reason.

An alternative explanation, applying to both SN 1993J and SN 1998S, is
based on the fact that both the \La\ and Mg~II lines are resonance
lines, and therefore optically very thick. This is in most cases true
also for \Ha\ \citep{CF94}.  If, in addition, the line-emitting region
is geometrically sufficiently thin, the Sobolev approximation will
break down and photons from the emitting shell may be absorbed
locally, giving rise to a tilted or double-peaked line
\citep[e.g.,][]{Rot52,KH74,BM87,CFT94}. Because both the hydrogen and
Mg~II lines are likely to be dominated by emission from the narrow CDS
between the reverse shock and the contact discontinuity (see \S
\ref{sec_origin}), this region may be too narrow for the Sobolev
approximation to be valid.

\cite{BM87} have discussed this case and find that ``M-shaped''
profiles result if three conditions are met: (1) The medium must be
optically thick in the line. (2) The shell must be thin enough for the
velocity {\it gradient} over the shell to be small compared to the
thermal velocity. (3) The macroscopic velocity must be larger than the
thermal velocity.
The reason for the high velocity peaks may be traced back to the fact
that the projected area contributing to a frequency interval $[\nu,
\nu+d\nu]$ in this limit is proportinal to $|\nu - \nu_0|$, where $
\nu_0$ is the rest ferquency of the line.

As we discuss above, the first of these conditions is true for \La\
and Mg~II, and in most cases also for \Ha. The second condition is
interesting. Lines coming from the ejecta, for example, are generally
formed over a fairly large velocity range \citep{CF94}, and are
therefore expected to have flat-topped profiles.  Lines coming from
the CDS, on the other hand, arise in a very narrow shell, no larger
than a few times $10^{12}$ cm \citep{CF94}. If we assume that
$v\propto r$ (which is not strictly valid in the CDS, but will provide
a conservative estimate), a velocity gradient equal to the thermal
velocity ($v_{\rm th}$) corresponds to a distance $\Delta r_{\rm th}
\approx v_{\rm th} \partial r / \partial v \approx R_{\rm CDS}\,
v_{\rm th}/V_{\rm CDS} = v_{\rm th}\, t \approx 10^{13} (t/100 ~{\rm
days})(T_{\rm e}/10^4~ {\rm K})^{1/2} A^{-1/2}$ cm. Here $A$ is the
atomic weight of the ion, $R_{\rm CDS}$ is the radius of the CDS, and
$T_{\rm e}$ the electron temperature in the CDS. Therefore, the velocity
gradient in the CDS is likely to be small enough for the the second
condition to be valid. This may, however, vary from case to case, as
well as over time. A ``microturbulence'' in the shell should have the
same effect. Condition (3) is fulfilled in all interesting
cases. Double-peaked line profiles can therefore be expected in cases
where the CDS is sufficiently narrow.

The fact that the \Ha\ line in SN 1998S evolves from a centrally
peaked line profile (before $\sim100$ days) to the double-peaked
profile at late epochs may be a result of the fact that the width of
the CDS at early epochs was large enough for the Sobolev case to apply
(note that $\Delta r_{\rm th}\propto t$). In this case, a nearly
constant velocity in the shell implies a parabolic line profile.

We therefore conclude that it is likely that {\it the double-peaked
profiles in SN 1998S and, to a less dramatic extent, SN 1993J arise as
result of the  extremely small thickness of the CDS, in
combination with a high optical depth}. Because the width of the CDS
is a function of the mass-loss rate and reverse-shock velocity
\citep[][]{CF94}, the line profiles provide a useful diagnostic of the
conditions behind the reverse shock.

The asymmetry of the red and blue wings of the lines may be caused by
either dust or bound-free absorption by the Balmer and Paschen
continua from the ejecta and circumstellar interaction region, as
discussed in Section \ref{sec_dust}. In the latter case, the larger
asymmetry for the Mg~II line compared to H$\alpha$ in SN 1993J would
then be a result of the larger optical depth in the Balmer
continuum. Dust absorption from the same region would have the same
effect. While bound-free absorption should decrease with decreasing
wavelength, dust absorption would increase. This could be checked by
comparison with the [N~II] $\wll 2139.7, 2143.5$
profile. Unfortunately, there are no observations of SN 1993J in the
far-UV region on day 670, when the Mg~II asymmetry is largest. Given
that there is strong evidence for dust emission, the dust
interpretation seems most likely. 

The origin of the central peak, if coming from the supernova rather
than the background H~II regions, is not clear. \citet{Ger00} suggested
emission from shocked circumstellar clumps, but it may also be
possible to explain it as an effect of the line formation process in
the cold, dense shell. 

We remark that this model does not naturally explain the polarization seen in
SN 1998S \citep{Leo00,Lifan01}. The origin of this may, however, be in
the density distribution of the ejecta inside of the interaction
region. A non-spherical cold dense shell, giving rise to the
polarization, would also be compatible with the
line profile.

\subsection{CNO Ratios in SN 1993J and SN 1998S}
\label{sec_cn}

The strong N~III] and N~IV] lines in both SN 1993J and SN 1998S
indicate nitrogen enrichment in the supernova ejecta and in the
circumstellar gas. To quantify this we make a nebular analysis similar
to that in \cite{FC89}. Assuming that the ionization zones of the
corresponding C and N ions coincide, which is reasonable
\citep[see][and references therein]{F84a,FC89}, one can calculate the
N~III/C~III and N~IV/C~IV ionic ratios as a function of temperature
and density. Because reasonable densities are considerably less than
the critical values for these ions ($\la 10^9 \cmc$), the density
is unimportant for the result. Further, because of the similar
excitation potentials of these lines, the temperature is only of minor
importance (see below). For the calculations below we use a
temperature of $2\times 10^4$ K and an electron density of $10^6
\cmc$. In the analysis we have used collision strengths and transition
probabilities from the compilation by \cite{PP95}.  For SN 1998S we
have used a reddening of $E_{B-V} = 0.23$ mag and for SN 1993J,
$E_{B-V} = 0.2$ mag \citep{R94,Leo00}.

For SN 1998S we find N~III/C~III $= 5.1 \pm 1.1$ and N~IV/C~IV $= 6.4
\pm 0.8$, where the errors have been estimated from the uncertainties
given in \S \ref{sec_nl}, and do not include systematic errors. From
the O~III] limit, we find a lower limit of N~III/O~III $> 1.4$.  In
Table \ref{tab2c}, we give the correction factors for these ionic
ratios for other temperatures and densities. We see that these never
exceed 20\%, and are therefore of the same order as the uncertainties
in the flux measurements.

For the N/C ratio we take a weighted average of the N~III/C~III and
N~IV/C~IV ratios, and obtain N/C $= 6.0 \pm 0.7$. In addition, there
are systematic errors from the uncertainty in the temperature and the
differences in the extents of the ionization zones. From Table
\ref{tab2c}, the former can be estimated to be at most 15\%. The
ionization structure depends on the ionizing source, which may be
either the X-rays from the shock wave, or possibly the flux in
connection to the breakout of the shock through the photosphere,
similar to the situation for the ring of SN 1987A. In that case
\cite{LF96} found that a detailed time-dependent photoionization model
resulted in N/C and N/O ratios which were $\sim35$\% lower than those
resulting from a simple nebular analysis along the lines in this
paper. As a realistic estimate of the N/C ratio we therefore take N/C
$=6.0 \pm 2$.

Similarly, we obtain for SN 1993J from the N~III]/C~III] ratio in
Table \ref{tab2a} the value N~III/C~III $= 12.4$. From the lower limit
of the N~IV] $\wl 1486$/C~IV $\wll 1548.9, 1550.8$ ratio of 1.5 we
obtain N~IV/C~IV $> 10.2$, which is compatible with the N~III/C~III
ratio. We therefore take N/C $=12.4$. The limit of N~III] $\wll
1746.8$--1754.0/O~III] $\wll 1660.8, 1666.2 > 1.0$ translates into
N~III/O~III $>0.8$, which we also take as a limit to the N/O ratio. As
we pointed out in \S \ref{sec_93j}, the feature at $\sim1640$~\AA\ is
more compatible in wavelength with He~II $\wl 1640$ than with the
O~III] line, and it is therefore likely that the N/O ratio is larger
than the lower limit given above.

Because of the complex deblending procedure in SN 1993J, as well as
its lower S/N, the errors in abundances are larger than for SN 1998S,
and we estimate that these may be $\pm 50\%$; thus, N/C $=12.4\pm 6$
and N/O $>0.8$.

To illustrate the level of carbon depletion in SN 1993J, we have
calculated a synthetic spectrum with solar ratios of C, N, and O. This
is shown as the dashed line in Figure \ref{fig4}, where we have kept
the N~III] flux at the same level. The flux of the C~III] line has
increased by a factor $\sim20$.

\subsection{Comparison with CNO Processing in Other Supernovae}

CNO abundance ratios have now been measured for a number of SNe.  The
SN~II-L 1979C was the first supernova shown to exhibit enhanced
nitrogen abundances; \cite{FB84} find N/C = $8 \pm 3$ and N/O $>
2$. The best-studied case to date is the ring of SN 1987A, from which
narrow emission lines of C~III-IV, N~III-V, and O~III-IV were
observed. Using a nebular analysis similar to the one presented 
in this paper,
\cite{FC89} found that N/C $= 7.8 \pm 4$ and N/O $=1.6\pm 0.8$.  Their
analysis was later improved using a self-consistent model of the ring
emission, yielding N/C = $5.0 \pm 2.0$ and N/O $= 1.1 \pm 0.4$
\citep{LF96}. The discrepancy between these abundance ratios gives a
reasonable estimate of the systematic errors due to differences in the
ionization zones, as well as the temperature of the gas. SN 1995N was
the first SN~IIn to have a C/N ratio determined, from a spectral
analysis of broad ejecta lines similar to that of SN 1979C and SN
1993J.  In Table \ref{tab3}, we summarize CNO ratios for SN 1979C, SN
1987A, SN 1993J, SN 1995N, and SN 1998S. Because of blending, the
broad-line determinations are affected by a larger uncertainty,
compared to values from the narrow circumstellar emission lines in SN
1987A and SN 1998S. This applies especially to SN 1995N.

From Table \ref{tab3} we see that in all cases where CNO abundances
have been measured, the N/C ratio is considerably larger than the
solar value, N/C $=0.25$ \citep{GS98}. The N/O ratio is more uncertain
due to the problems with O~III] $\wl 1664$. The only well-determined
case is SN 1987A, where N/O $\approx 1.1$, again much larger than the
solar value N/O $=0.12$.

All supernova progenitors in Table \ref{tab3} are believed to have
undergone extensive mass loss prior to the explosion. Models of SN
1987A suggest that $\sim6 \Msun$, out of the original $\sim14 \Msun$
hydrogen envelope mass, were lost before the explosion
\citep[e.g.,][]{SN90}.  This may be characteristic for Type II-P
progenitors, although one should keep in mind that the SN 1987A
progenitor was not typical for a Type II-P, and that the progenitor
mass was relatively high. For the other Type II subclasses,
\citet{NIS95} have proposed a binary scenario, where the SNe~II-L and
SNe~IIn have lost a moderate amount of the hydrogen envelope, while
SNe~IIb have an envelope mass $\la 1 \Msun$. Type Ib and Ic SNe have
lost all of their hydrogen envelope, and in the case of SNe~Ic also
the helium mantle.  As we discussed in \S \ref{sec_origin}, for SN
1993J, SN 1995N, and SN 1998S there is strong evidence that the
progenitors had lost most of their hydrogen envelopes before the
explosion.

With this background, it is tempting to see the nitrogen enrichment in
these SNe as being a result of this mass loss.  A more quantitative
comparison is, however, not straightforward.  Stellar evolutionary
models of massive stars, including effects of mass loss, rotation, and
binarity, have been calculated by several groups. In particular,
rotation can have a large effect on the CNO abundances by increasing
the mixing from the CNO burning region already in the main-sequence
phase \citep{MM00,HL00}. Binary mass exchange produces a result which,
unfortunately, is not easily distinguished from that of rotation
\citep{WL99}.

The nitrogen enrichment in SN 1987A has put considerable constraints
on progenitor models \citep[for reviews see][]{Pod92,WHW02}. Here the
problem is to explain the blue progenitor, the red-to-blue supergiant
ratio in the LMC, and the CNO ratios in a consistent model. To fulfill
all these constraints has turned out to be difficult in single-star
scenarios. For example, \cite{SNK88} had to artificially invoke 
complete mixing of the hydrogen envelope.

\citet{HM89}, \citet{PJP90}, and \citet{Ray93} have argued that binary
models, in particular merger models, should be more successful in
these respects. The nitrogen enrichment is then a natural consequence
of the merger or accretion process, which adds both angular momentum
and energy; this may in turn induce large-scale mixing of the
envelope. In addition, the triple-ring structure may be more naturally
explained in this kind of scenario. A problem may be that such models
require some fine tuning of the precise epoch when the merger takes
place \cite[see, e.g.,][]{WHW02}.

Of the remaining SNe, the N/C ratio of SN 1993J may be the easiest one
to explain. Light-curve calculations \citep{Nom93,WEWP94} and spectral
models \citep{PCM95,HF96} have shown that the hydrogen envelope of SN
1993J must have had a mass of less than $1 \Msun$ at the time of the
explosion. For this to occur without too much fine tuning, a binary
model is most natural. Both of these properties make mixing of
CNO-processed products to the surface natural.

Nitrogen enrichment has also been found in older supernova remnants,
most notably in Cas A. \cite{CK78} discovered an overabundance of
nitrogen by an order of magnitude in the quasi-stationary flocculi,
presumably originating from the clumpy wind of the
progenitor. \cite{L78} explained this enrichment as a result of CNO
burning in combination with mass loss from a progenitor of
15--25~$\Msun$. Moreover, \cite{FB91} and \cite{Fes01} identified {\it
fast-moving knots} rich in both hydrogen and nitrogen. Some of these
were found outside the main blast wave. These must have been expelled
from the photosphere of the progenitor at the time of the explosion,
showing that the progenitor had at least a thin nitrogen-enriched
hydrogen envelope. Based on these observations \cite{CO03} have
suggested that the progenitor of Cas A was a Type IIb or IIn
supernova. The broad wings seen in H$\alpha$ for both SN 1998S and SN
1995N \citep{FC02} may have originated in comparable high-velocity
clumps, which have penetrated the supernova blast wave. The narrow
component may be similar to the quasi-stationary flocculi. Finally,
for Cas A we have already noted the possibility of a similar scenario
for the dust formation as for SN 1998S.

\section{CONCLUSIONS}

The most important result of the observations presented here is the
strong indication of nitrogen enrichment in both SN 1993J and SN
1998S. This, together with similar evidence from other SNe~II, puts
strong constraints on their progenitors. The CNO enrichment also gives
independent support for mass loss from the progenitors. The line
profiles and the presence of high-ionization lines are direct
indicators of the ionizing effects from the reverse shock wave, in
qualitative agreement with that expected from models of circumstellar
interaction. The double-peaked nature of the line profiles is
interpreted as a result of the emission from the cool dense shell
behind the reverse shock, rather than a torus, and can serve as an
important diagnostic of the former. The quantitative features of this,
however, remain to be demonstrated.  From the asymmetry of the line
profiles in SN 1998S we find additional support for dust formation in
the ejecta.

\acknowledgments

We are grateful to the referee for constructive comments. Financial
support for this work was provided to the Stockholm group by the
Swedish Space Board and Swedish Research Council, and to R.A.C. by NSF
grant AST-0307366.  A.V.F. is grateful for NSF grants AST-9987438 and
AST-0307894, as well as for NASA grant AR-9529 from the Space
Telescope Science Institute (STScI), which is operated by AURA, Inc.,
under NASA contract NAS 5-26555. The {\it HST} SINS collaboration has
most recently been supported by NASA through grants GO-9114 and
GO-9428 from STScI.  Some of the data presented herein were obtained
at the W. M. Keck Observatory, which is operated as a scientific
partnership among the California Institute of Technology, the
University of California, and NASA; the Observatory was made possible
by the generous financial support of the W. M. Keck Foundation.  We
thank the staffs at the {\it HST}, FLWO, Lick, and Keck Observatories
for their overall assistance.



\clearpage


\clearpage

\clearpage

\clearpage

\clearpage

\clearpage

\clearpage

\clearpage

\clearpage

\clearpage

\clearpage

\clearpage

\clearpage






\clearpage

\begin{deluxetable}{lrlcrr}
\tabletypesize{\scriptsize}
\tablecaption{Log of {\it HST} FOS and STIS observations of SN 1993J\label{tab1a}}
\tablewidth{0pt}
\tablehead{
\colhead{UT Date}  & 
\colhead{Epoch\tablenotemark{a}}  & \colhead{Grating\tablenotemark{b}}  &
\colhead{Dispersion}  &
\colhead{Exposure} &\colhead{Range} \\
\colhead{}     &\colhead{(days)} &\colhead{} &
\colhead{(\AA\ pixel$^{-1}$)}&\colhead{(s)}&\colhead{(\AA)}}
\startdata 
26 January 1995&670 & G270H&2.05&3600&2221--3301  \\
               &    & G400H&3.00&2600 &3240--4822  \\
23 February 1996&1063 &  G160L&6.64&8200 &1140--2508  \\
                &     &  G270H&2.05&4800 &2221--3301  \\
24 January 1997&1399 &  G160L&6.64&22200 &1140--2508  \\
               &     &  G270H&2.05&6400 &2221--3301  \\
21 February 1998 &1792 &  G230L&1.58&9300 &1568--3184  \\
17 April 1998 &1847  &  G140L&0.60&20200 &1140--1730  \\
18 April 2000 & 2579 &  G140L&0.60&19500&1140--1730  \\
24 April 2000 & 2585 &  G230L&1.58&3000 &1570--3180  \\
\enddata
\tablenotetext{a}{Based on an assumed explosion date of 27.5 March
1993 UT.}
\tablenotetext{b}{FOS, 1995--1997; STIS, 1998--2000.}
\end{deluxetable}

\begin{deluxetable}{lrlcrr}
\tabletypesize{\scriptsize}
\tablecaption{Log of {\it HST} STIS observations of SN 1998S\label{tab1b}}
\tablewidth{0pt}
\tablehead{
\colhead{UT Date}  & 
\colhead{Epoch\tablenotemark{a}}  & \colhead{Grating}  &
\colhead{Dispersion}  &
\colhead{Exposure} &\colhead{Range} \\
\colhead{}     &\colhead{(days)} &\colhead{} &
\colhead{(\AA\ pixel$^{-1}$)}&\colhead{(s)}&\colhead{(\AA)}}
\startdata 
16 March 1998&14&G140L&0.60&2160&1140--1730  \\
             &  &G230L&1.58&3920&1570--3180 \\
             &  &G430L&2.73& 180 &2900--5700  \\
30 March 1998&28&G140L&0.60&6860&1140--1730  \\
             &  &G230L&1.58&2400&1570--3180 \\
             &  &G430L&2.73& 300 &2900--5700  \\
13 May 1998&72&G140L&2.05&6860&1140--1730  \\
             &  &G230L&2.05&5200&1570--3180  \\
             &  &G430L&3.00& 300 &2900--5700 \\
26 Oct 1998&238&G140L&2.05&6730&1140--1730  \\
             &  &G230L&2.05&5000&1570--3180  \\
             &  &G430L&3.00& 700 &3240--4822  \\
30 Jun 1999&485&G140L&2.05&5500&1140--1730  \\
             &  &G230L&2.05&11600&1570--3180  \\
\enddata
\tablenotetext{a}{From the date of discovery on 2 March 1998 UT (see text).}
\end{deluxetable}

\begin{deluxetable}{lrcccccccrrl}
\rotate
\tabletypesize{\scriptsize}
\tablecaption{Log of ground-based observations of SN 1998S\label{tab1c}}
\tablewidth{0pt}
\tablehead{
\colhead{UT Date}  & 
\colhead{Epoch\tablenotemark{a}}  &\colhead{Instrument}&\colhead{Range}&\colhead{Resolution}  &
\colhead{P.A.}  &
\colhead{Parall. A.}&\colhead{airmass}&\colhead{Standard star}&\colhead{seeing}&\colhead{slit}&\colhead{Exposure}  \\
\colhead{}     &\colhead{(day)} & &\colhead{(\AA)} &
\colhead{(\AA)}&\colhead{(deg)}&\colhead{(deg)}&\colhead{}&\colhead{}&\colhead{($''$)}&\colhead{($''$)}&\colhead{(s)}}
\startdata 
1998-03-31&29&FLWO &3720--7521&7.0&90&65&1.12 & Feige 34 & 2-3 & 3 & 3 $\times$300 \\
1998-05-16&75&FLWO &3720--7540&7.0&90&177&1.04 & Feige 34 & 1-2 & 3 & 660 \\
1998-06-18&108&Lick&3300--5400&8&66&69&1.83&Feige 34&1.5&2&2 $\times$ 300\\
1998-06-18&108&Lick&5100--10400&12&66&69&1.83&HD 84937&1.5&2&2 $\times$ 300\\
1998-07-17&137&Lick&3300-- 5400& 8&71&71&1.74&BD +284211&2.0&2&900\\
1998-07-17&137&Lick&5200--10400&12&71&71&1.74&BD +262606&2.0&2&900\\
1998-07-23&143&Lick&4300--7050&10&73&72&1.68&BD +262606&1.5&2&900\\
1998-07-23&143&Lick&3280--5420&6&73&72&1.78&BD +284211&1.5&2&720\\
1998-07-23&143&Lick&6050--8050&7&73&72&1.78&BD +262606&1.5&2&720\\
1998-11-14&257 &FLWO& 3720--7521 & 7.0 &90& 81 & 1.19 & Feige 34 & 3 & 3 & 2 $\times$ 1200 \\
1998-11-15&258 &FLWO& 3720--7521 & 7.0 &90& 83 & 1.22 & Feige 34 & 2-3 & 3 & 3 $\times$ 1200  \\
1998-11-25&268&Lick&5200--10200&12&88&87&1.25&BD +174708&2.2&2&1272+2000\\
1999-01-06&310&Keck2&4360--6860&6&175&184&1.13&HD19445&1.2&&450\\
1999-01-10&314&Lick&3350--5500&8&180&38&1.02&BD +284211&1.5&2&2 $\times$ 1800\\
1999-01-10&314&Lick&5200--10500&12&180&38&1.02&BD +74408&1.5&2&2 $\times$ 1800\\
1999-03-12&375&Lick&3276--5400&8&185&115&1.06&Feige 34&1.8&2&1800\\
1999-03-12&375&Lick&5250--10500&12&185&115&1.06&HD 19445&1.8&2&1800\\
1999-05-07 & 431 & FLWO &3720--7521 & 7.0 & 0 & 175 & 1.04 & Feige 66 & 1-2 & 3 & 3 $\times$ 1200 \\
1999-05-16 & 440 & FLWO &3720--7540 & 7.0 & 90 &  127 & 1.12 & Feige 34 & 1-2 & 3 & 4 $\times$ 1200 \\
1999-07-09&494&Lick&3290--5400&8&70&71&1.7&Feige 34&2.3&3&2 $\times$ 1500\\
1999-07-09&494&Lick&5156--10550&12&70&71&1.7&BD +262606&2.3&3&2 $\times$ 1500\\
1999-12-05&643&Keck2&4000--10000&15 & 50 & 51 &1.2 & HD 84937&1.8 & 1.0 &900\\
1999-12-15&653&Keck2&5644--6935&2.5 & 65 & 65 &1.37 & HD 84937&1 & 1.0 &300+600\\
2004-01-18 &2148&MMT&3270--8460&8.0&130& 122 & 1.07 &Feige 34/HD 84937 &1.5&2&2 $\times$ 1800 \\

\enddata
\tablenotetext{a}{From the date of discovery on 2 March 1998 UT (see text).}
\end{deluxetable}

\begin{deluxetable}{lcr}
\tabletypesize{\scriptsize}
\tablecaption{Line fluxes, SN 1993J \label{tab2a}}
\tablewidth{0pt}
\tablehead{
\colhead{Line}     & \colhead{Relative flux\tablenotemark{a}} \\
}
\startdata
 ~He~II               1640.4           &   0.50        \\ 
 ~O~III]               1660.8, 1666.1   &   $0.13$           \\ 
 ~N~III]           1746.8--1753.4      &   1.00          \\
 ~Ne~III]           1814.6              &   1.00          \\
 ~Si~III]          1882.7, 1892.0       &   0.12         \\
 ~C~III]           1906.7, 1909.6       &   0.63         \\
\enddata
\tablenotetext{a}{Fluxes relative to N~III].}
\end{deluxetable}

\
\begin{deluxetable}{lccrrrr}
\tabletypesize{\scriptsize}
\tablecaption{Fluxes of the narrow lines in SN 1998S\tablenotemark{a}\label{tab2}}
\tablewidth{0pt}
\tablehead{
\colhead{Line}  & 
\colhead{Rest }  & \colhead{Observed}  &
\colhead{Day 28} &\colhead{Day 72}&\colhead{Day 238} &\colhead{Day 485} \\
\colhead{}     &\colhead{wavelength} &\colhead{wavelength}
&\colhead{30 March}& \colhead{13 May}& \colhead{26 October}& \colhead{30 June} \\
\colhead{}     &\colhead{(\AA)} &\colhead{(\AA)} &
\colhead{1998} & \colhead{1998}& \colhead{1998}& \colhead{1999}  
}
\startdata
 ~N~V     &1238.8, 1242.8&  1238.5 1240.6 & 3.0:& 1.10 & 1.6  & 1.4 \\
 ~Si~IV     &1393.8&    1394.0     &19.1 & 0.88     &         &         \\
 ~Si~IV     & 1402.8&  1403.1    &26.1 & 1.27     &         &         \\
 ~N~IV]           &1483.3--1487.9 & 1486.3  & 5.8 & 2.49 &1.40: &2.16  \\
 ~C~IV             &  1548.9, 1550.8   &  1549.2  &          & 2.66 & 0.91 & 1.03   \\
 ~N~III]           &1746.8--1753.4    & 1751.1  &  18.2:        & 2.47 & &           \\
 ~C~III]   &1906.7, 1909.6& 1909.3  & 36.5:  & 3.77     & 1.43       &  0.41:    \\
 ~C~II]   &2322.7--2328.8& 2325.0  & 1.7  & 1.3     &       &      \\
\enddata
\tablenotetext{a}{Observed fluxes are in units of $10^{-15} \ergs {\rm ~cm}^{-2}$.}
\end{deluxetable}

\begin{deluxetable}{lcccc}
\tabletypesize{\scriptsize}
\tablecaption{Temperature and density correction factors \label{tab2c}}
\tablewidth{0pt}
\tablehead{
\colhead{Temperature}  & \colhead{Density}  & 
\colhead{N~III/C~III}  & \colhead{N~IV/C~IV} & \colhead{N~III/O~III} \\
\colhead{(K)}     &\colhead{(cm$^{-3}$)} & \colhead{}& \colhead{} \\
}
\startdata
10,000&$10^6$&1.16&1.09&0.70 \\
15,000&$10^6$&1.04&1.02&0.88 \\
20,000&$10^4$&1.04&1.00&1.06 \\
20,000&$10^6$&1.00&1.00&1.00 \\
20,000&$10^8$&0.96&1.00&0.93 \\
30,000&$10^6$&0.95&1.00&1.16 \\
\enddata
\end{deluxetable}

\begin{deluxetable}{lccrrl}
\tabletypesize{\scriptsize}
\tablecaption{Summary of CNO abundances in SNe~II\label{tab3}}
\tablewidth{0pt}
\tablehead{
\colhead{Supernova}  & 
\colhead{Type}  & \colhead{environment}& \colhead{N/C}&
\colhead{N/O}&\colhead{Notes}  \\
}
\startdata
 ~SN 1979C    & II-L&ejecta&  8&$> 2$&\\
 ~SN 1987A     &II-P&circumstellar&  7.8 & 1.6&nebular analysis \\
              &   &                 &  5.0 & 1.1&photoionization model\\
 ~SN 1993J  &IIb&ejecta&12.4&$>0.8$&\\
 ~SN 1995N  &IIn&ejecta&3.8&0.2&uncertain\\
 ~SN 1998S  &IIn &circumstellar&6.0&$> 1.4$& \\
\enddata
\end{deluxetable}


\end{document}